\documentclass[12pt]{emulateapj}
\slugcomment{{\em Submitted to ApJ.} -- \today}

\usepackage{epsfig}
\usepackage{amsmath}
\usepackage{amssymb}
\usepackage{natbib}
\usepackage{graphicx}

\shorttitle{The Two Phases of Galaxy Formation}
\shortauthors{Oser et al.} 

\begin{document}

\title{The Two Phases of Galaxy Formation}
%\title{}

\author{Ludwig Oser$^1$$^2$$^4$, Jeremiah P. Ostriker$^3$, Thorsten
  Naab$^1$$^4$, Peter H. Johansson$^1$, Andreas Burkert$^1$} 
\affil{$^1$ Universit\"ats-Sternwarte M\"unchen, Scheinerstr.\ 1,
  D-81679 M\"unchen, Germany; \texttt{oser@usm.lmu.de} \\ 
$^2$ Max-Planck-Institut f\"ur extraterrestrische Physik, Giessenbachstrasse,
 D-85748 Garching, Germany\\
$^3$ Department of Astrophysical Sciences, Princeton University,
  Princeton, NJ 08544, USA\\ 
$^4$ Max-Planck-Institut f\"ur
  Astrophysik, Karl-Schwarzschild-Strasse 1, 85741 Garching, Germany} 

\begin{abstract}
Cosmological simulations of galaxy formation appear to show a
'two-phase' character with a rapid early phase at z$\gtrsim$2 during
which 'in-situ' stars are formed within the galaxy from infalling
cold gas followed by an extended phase since z$\lesssim$3 during which
'ex-situ' stars are primarily accreted. In the latter phase massive systems grow
considerably in mass and radius by accretion of smaller satellite
stellar systems formed at quite early times ($z > 3$) outside of the
virial radius of the forming central galaxy. These tentative conclusions are
obtained from high resolution re-simulations of 39 individual galaxies
in a full cosmological context with present-day virial halo masses
ranging from $7 \times 10^{11} M_{\odot}h^{-1}$ $ \lesssim
M_{\mathrm{vir}} \lesssim 2.7 \times 10^{13} M_{\odot}h^{-1}$ (h=0.72)
and central galaxy masses between $4.5 \times 10^{10} M_{\odot}h^{-1}
\lesssim M_* \lesssim 3.6 \times 10^{11} M_{\odot}h^{-1}$. The
simulations include the effects of a uniform UV background, radiative
cooling, star formation and energetic feedback from SNII. The
importance of stellar accretion increases with galaxy mass and towards
lower redshift. In our simulations lower mass galaxies ($M_* \lesssim 9 \times
10^{10}M_{\odot}h^{-1}$) accrete about 60 per cent of their
present-day stellar mass. High mass galaxy ($M_* \gtrsim 1.7 \times
10^{11}M_{\odot}h^{-1}$) assembly is dominated by accretion and merging
with about 80 per cent of the stars added by the present-day. 
In general the simulated galaxies approximately double their mass since z=1.
For massive systems this mass growth is not accompanied by significant star formation.
The majority of the in-situ created stars is formed
at $z>2$, primarily out of cold gas flows.
We recover the observational result of 'archaeological
downsizing', where the most massive galaxies harbor the oldest
stars. We find that this is not in contradiction with hierarchical
structure formation. Most stars in the massive galaxies are
formed early on in smaller structures, the galaxies
themselves are assembled late.  
\end{abstract}


\section{Introduction}
\label{introduction}

Our understanding of galaxy formation has made great advances in the last two decades 
driven - primarily - by technological progress. Both ground and sky based measurements
 have allowed direct observation of various phases of galaxy formation and evolution 
over cosmic time with some detailed information now available at redshifts $z > 2$ 
(e.g. \citealp{1999ApJ...519....1S,2001ApJ...554..981P,2006Natur.442..786G,2006ApJ...645.1062F,
2007MNRAS.382..109T,2008ApJ...682..896K,2008ApJ...677L...5V,2009ApJ...701.1765M,2009ApJ...706.1364F}). 
Simultaneously with a quite definite cosmological model ($\Lambda CDM$, e.g. \citealp{2007ApJS..170..377S}, 
\citealp{2010arXiv1001.4538K}) having gained wide acceptance, we can, with increasing accuracy, simulate the 
evolution of galaxies from realistic initial conditions, with numerical resolution (in mass, space, and time) 
and physical modeling approaching the necessary degree of refinement (e.g. \citealp{2003ApJ...596...47S,
2003MNRAS.339..289S,2005MNRAS.364.1105S,2005ApJ...627..608N,2007ApJ...658..710N,2007MNRAS.374.1479G,
2009arXiv0909.4167P,2009MNRAS.396..696S,2010MNRAS.402.1599S,2010arXiv1004.0005A,2010MNRAS.402.1536S}) 

The overall results are reassuring, with simulations and observations agreeing - in gross outline - 
as to the time evolution of star/galaxy formation (e.g. \citealp{2006ApJ...653..881N,2009arXiv0909.5196S})
as well as the global attributes of the galaxies such as luminosity distribution and spatial organization
(e.g. \citealp{1999ApJ...514....1C,1999MNRAS.303..188K,1999MNRAS.310.1087S,2005Natur.435..629S,
2009MNRAS.396.2332K,2010arXiv1006.0106G}). Understanding the development of the internal structures 
of galaxies has been far more difficult to achieve with respect to the galactic stellar mass fractions 
(e.g. \citealp{2009MNRAS.396.2332K,2010MNRAS.404.1111G}) as well as kinematics and morphologies 
(e.g. \citealp{2003ApJ...597...21A,2010Natur.463..203G,2010ApJ...709..218F}). 

The terms with which we might usefully describe 
such development are still controversial (e.g. \citealp{2003ApJ...590..619M,2007ApJ...658..710N,
2007MNRAS.374.1479G,2009arXiv0909.4167P}). 
In a hierarchically organized universe it has been natural to focus on overdense 'lumps' 
of dark matter gas or stars and to follow the merger history of these lumps. A recent paper by 
\citet{2009arXiv0906.5357H} shows how useful this picture can be. But this is not the only description 
of galaxy formation. For example \citet{2005MNRAS.363....2K,2009MNRAS.395..160K} and 
\citet{2009ApJ...703..785D} have focused on how convergent cold streams of gas lead to early 
star bursts and the formation of the cores of 
massive galaxies. \citet{2007ApJ...658..710N,2009ApJ...699L.178N}, \citet{2009ApJ...692L...1J} and 
others have used high resolution hydro simulations to explore this phase in greater detail 
(see also \citet{2005MNRAS.359...93M} for the accretion histories of stellar halos of disk galaxies).

One fundamental and useful distinction is to examine if a given star in the final galaxy was made (from gas) 
close to the center of the final system or, alternatively, near the
center of some other, distant system and accreted in stellar form to
the final galaxy. This distinction is useful, e.g. for understanding
the size evolution of massive galaxies
\citep{2006ApJ...648L..21K,2007ApJ...658..710N,2009ApJ...699L.178N, 
2009ApJ...697.1290B,2009ApJ...706L..86N,2010MNRAS.401.1099H,2010ApJ...709..218F}. In the simulations 
presented here we find that most stellar particles in massive galaxies are formed at high redshift 
either far inside the virial radius ($\lesssim 3 \rm kpc$) near the
forming galaxy center or, alternatively in small 
systems outside the virial radius of the galaxy at a given cosmic time. We characterize the first 
category of stars as made 'in-situ' and the second as accreted or formed 'ex-situ'. In-situ stars are made 
(by definition) near to the galactic center over an extended time period. They are made from dissipative 
gas and, for massive systems, probably have relatively high metalicity \citep{2010ApJ...721..738Z}. 
The peak rate of star formation for this category may be relatively early and in fact is very early 
($z \approx 4$) for the most massive systems.

On the contrary, the accreted stars are typically made at quite early times as well, outside the 
virial radius, but added to the parent galaxy late in its evolution. They are added typically 
at radii larger than the effective radius, $r> r_{\mathrm{eff}}$, and
are expected to be metal poor, 
since they originated in lower mass, lower metallicity systems \citep{2009ApJ...699L.178N}. 
The ex-situ stars accrete via an energetically conservative process and their final binding energy 
is transferred to other phases (gas, stars, and dark matter) rather than simply radiated away 
\citep{2009ApJ...697L..38J}.

This alternative way of envisioning galaxy formation has many corollaries and makes many observed facts 
easier to understand. In massive systems we expect considerable growth in mass and radius at late times 
but little star formation, with the late forming stellar envelopes comprised of stars which are typically 
older than the stars in the bulk of the galaxy. Further we find systematic trends with galaxy mass. 
As one considers systems of lower mass, the in-situ component becomes increasingly dominant and 
the period of in-situ star formation is stretched out from being a small fraction of the Hubble time to a 
large fraction thereof.

The paper is organized as follows. In section \ref{sim} we describe our simulations in detail, as 
there will be follow-up papers using this set of simulations. In addition we here discuss the
conversion efficiency of gas into stars for our simulated galaxies. 
In section \ref{twophases} we examine the dependence of the ratio of in-situ formed to accreted stars 
on the galaxy stellar mass along with 
its implications. We go on to analyze the half-mass radii of the different stellar components of our 
simulated galaxies in section \ref{galsizes}. Finally, in section \ref{conc} we summarize our 
findings.

\begin{table*}
\caption{Central Galaxies}
\centering
%\begin{tabular}{c | c | c | c | c | c}
\begin{tabular}{c | c | c | c | c | c | c | c | c | c | c | c | c }

\hline\hline
%ID & $m_{200}$ & $m_{*,tot}$ & $m_{*,ins}$ & $m_{*,acc}$ & $m_{gas}$  \\
ID & $m_{200}$\footnotemark[1] & $r_{200}$\footnotemark[2] & $m_{*}$\footnotemark[3] & $m_{gas}$\footnotemark[4] & 
$m_{ins}/m_*$\footnotemark[5] & $t_*$\footnotemark[6]& $t_{ins}$\footnotemark[7]& $t_{acc}$\footnotemark[8]& 
$t_{50}$\footnotemark[9] & $n_{gas}$\footnotemark[10] & $n_*$\footnotemark[11] & $n_{halo}$\footnotemark[12] \\
\hline
0040 &     2676 &     486  &     36.0 &     4.13 &    0.231 &     10.8 &     9.90 &     11.1 &     2.73 &   579933 &   440633 &  2096930 \\
0069 &     1775 &     424  &     35.6 &     3.13 &    0.218 &     10.8 &     8.66 &     11.4 &     6.37 &   354378 &   306742 &  1378352 \\
0089 &     1064 &     358  &     37.7 &     2.58 &    0.163 &     11.0 &     9.91 &     11.2 &     4.75 &   214528 &   182465 &   826895 \\
0094 &     1004 &     351  &     34.5 &     3.46 &    0.258 &     10.9 &     9.10 &     11.6 &     7.67 &   210596 &   164402 &   780411 \\
0125 &     917  &     340  &     31.2 &     2.94 &    0.224 &     11.1 &     9.59 &     11.6 &     8.31 &   200865 &   146889 &   716832 \\
0162 &     630  &     300  &     26.2 &     2.64 &    0.129 &     10.8 &     8.49 &     11.2 &     2.58 &   134454 &   106554 &   494315 \\
0163 &     689  &     309  &     25.3 &     1.73 &    0.150 &     10.5 &     9.11 &     10.8 &     4.75 &   139297 &   119486 &   536504 \\
0175 &     699  &     311  &     26.5 &     1.29 &    0.270 &     11.3 &     9.74 &     11.8 &     9.56 &   127745 &   117170 &   530274 \\
0190 &     511  &     280  &     22.7 &     1.71 &    0.146 &     10.3 &     8.39 &     10.6 &     3.81 &   103075 &    98844 &   405894 \\
0204 &     538  &     285  &     19.3 &     1.42 &    0.156 &     10.8 &     8.77 &     11.2 &     8.31 &   102722 &    99548 &   419003 \\
0209 &     595  &     295  &     14.4 &    0.656 &    0.337 &     10.9 &     9.71 &     11.5 &     9.26 &   118459 &    97601 &   457580 \\
0215 &     505  &     279  &     19.9 &     1.14 &    0.352 &     11.0 &     9.93 &     11.5 &     8.15 &   100251 &    87072 &   391385 \\
0224 &     478  &     274  &     17.9 &     2.06 &    0.200 &     10.3 &     7.69 &     11.0 &     6.20 &    89336 &    91799 &   373489 \\
\tableline
0259 &     437  &     266  &     14.3 &     1.23 &    0.262 &     10.9 &     8.98 &     11.6 &     9.72 &    83945 &    81751 &   341491 \\
0300 &     365  &     250  &     13.4 &     1.63 &    0.201 &     10.4 &     8.64 &     10.8 &     5.88 &    72180 &    64276 &   283964 \\
0329 &     350  &     247  &     15.4 &    0.696 &    0.341 &     10.9 &     9.55 &     11.6 &     9.10 &    65296 &    63583 &   270652 \\
0380 &     328  &     242  &     12.3 &    0.634 &    0.491 &     10.9 &     10.0 &     11.8 &     10.6 &    58842 &    56464 &   249316 \\
0408 &     253  &     221  &     12.8 &     1.90 &    0.300 &     10.1 &     7.09 &     11.3 &     8.31 &    49561 &    50348 &   200794 \\
0443 &     268  &     226  &     16.6 &     1.91 &    0.277 &     10.3 &     6.55 &     11.7 &     8.31 &    50289 &    52800 &   210493 \\
0501 &     230  &     215  &     11.7 &    0.93 &    0.361 &     10.8 &     10.0 &     11.2 &     8.79 &    48521 &    40463 &   181178 \\
0549 &     216  &     210  &     8.38 &    0.450 &    0.262 &     10.7 &     8.71 &     11.4 &     9.41 &    39034 &    39605 &   165346 \\
0616 &     189  &     201  &     9.38 &    0.455 &    0.367 &     10.6 &     9.88 &     11.0 &     9.72 &    34520 &    37188 &   147962 \\
0664 &     179  &     197  &     7.48 &    0.558 &    0.343 &     10.7 &     9.06 &     11.6 &     9.41 &    34393 &    30862 &   138039 \\
0721 &     147  &     185  &     9.63 &    0.629 &    0.536 &     8.88 &     7.07 &     11.0 &     6.69 &    22910 &    34776 &   116680 \\
0763 &     150  &     186  &     9.85 &    0.177 &    0.197 &     10.4 &     9.19 &     10.8 &     6.37 &    25283 &    34151 &   119180 \\
0858 &     139  &     181  &     10.3 &    0.790 &    0.355 &     8.92 &     5.49 &     10.8 &     6.69 &    21022 &    33759 &   110365 \\
\tableline
0908 &     125  &     175  &     9.67 &    0.708 &    0.458 &     8.84 &     6.55 &     10.8 &     7.50 &    19927 &    33080 &   102025 \\
0948 &     121  &     173  &     6.64 &    0.442 &    0.308 &     10.6 &     9.38 &     11.2 &     9.56 &    22627 &    23147 &    94475 \\
0959 &     120  &     173  &     6.05 &    0.399 &    0.371 &     10.1 &     9.46 &     10.5 &     9.41 &    23591 &    23027 &    94670 \\
0977 &     94.4 &     159  &     4.55 &    0.464 &    0.496 &     9.10 &     7.21 &     11.0 &     8.63 &    16592 &    20916 &    75143 \\
1017 &     106  &     166  &     6.39 &    0.736 &    0.584 &     10.0 &     8.92 &     11.5 &     9.87 &    21049 &    20634 &    83999 \\
1061 &     103  &     164  &     5.18 &    0.439 &    0.335 &     9.98 &     8.72 &     10.6 &     8.15 &    19196 &    20400 &    81076 \\
1071 &     106  &     166  &     7.79 &    0.610 &    0.317 &     9.66 &     7.06 &     10.9 &     8.15 &    18696 &    24045 &    84818 \\
1091 &     112  &     169  &     7.53 &    0.416 &    0.280 &     9.24 &     5.37 &     10.7 &     6.20 &    18487 &    26210 &    89119 \\
1167 &     93.0 &     159  &     7.37 &    0.659 &    0.331 &     9.32 &     5.88 &     11.0 &     6.85 &    15966 &    22371 &    75141 \\
1192 &     78.0 &     150  &     4.36 &    0.157 &    0.442 &     10.4 &     9.54 &     11.0 &     9.56 &    13041 &    15792 &    60404 \\
1196 &     95.4 &     160  &     7.73 &    0.99 &    0.490 &     9.23 &     6.96 &     11.4 &     7.67 &    16839 &    20987 &    75883 \\
1646 &     71.3 &     145  &     5.38 &    0.509 &    0.480 &     8.90 &     6.23 &     11.4 &     8.31 &    11143 &    16557 &    56264 \\
1859 &     70.0 &     144  &     4.52 &    0.340 &    0.429 &     9.82 &     7.86 &     11.3 &     9.56 &    12355 &    16458 &    56488 
%2283 &     49.1 &     128  &     3.22 &    0.126 &    0.488 &     8.61 &     7.43 &     9.73 &     6.69 &     4151 &     9514 &    27555 \\
%2665 &     38.8 &     119  &     3.17 &    0.174 &    0.433 &     9.36 &     8.18 &     10.3 &     7.67 &     5784 &    10038 &    31227 \\
%3431 &     20.9 &     96.5 &     1.87 &    0.221 &    0.730 &     8.66 &     7.80 &     11.0 &     9.10 &     3261 &     5379 &    16932 \\
%3852 &     30.5 &     109  &     2.64 &   0.098 &    0.432 &     9.18 &     7.73 &     10.3 &     6.53 &     4632 &     8541 &    25117 \\
%4323 &     21.2 &     96.9 &     2.58 &   0.0535 &    0.454 &     8.27 &     6.21 &     9.98 &     3.03 &     3495 &     7038 &    20804 \\
%5014 &     23.4 &     100  &     2.26 &    0.198 &    0.475 &     8.21 &     6.99 &     9.32 &     7.67 &     3810 &     6581 &    19483 \\
%6782 &     15.8 &     87.9 &     1.95 &   0.0857 &    0.353 &     8.89 &     7.23 &     9.80 &     3.96 &     2621 &     5466 &    14021 
\end{tabular}

\tablecomments{all masses in units of $10^{10}h^{-1}M_{\odot}$, timescales in Gyr. \\
\footnotemark[1]virial mass, \footnotemark[2]virial radius in $\rm kpc/h$, \footnotemark[3]{stellar mass inside $r_{10}$} , \footnotemark[4]{gas mass inside $r_{10}$}, 
\footnotemark[5]{ratio of in-situ to ex-situ created stars}, \footnotemark[6]{mean stellar age} , 
\footnotemark[7]{mean stellar age of in-situ created stars}, \footnotemark[8]{mean stellar age of ex-situ created stars}, \footnotemark[9]{lookback time where 50 per cent of the final stellar mass is in place}, \footnotemark[10]number of gas particles inside $r_{200}$, 
\footnotemark[11]number of star particles inside $r_{200}$, \footnotemark[12]total number of particles inside $r_{200}$. \\
 The horizontal bars indicate the separation into small, intermediate and high mass galaxies used throughout this paper}

\label{tab:ins-acc}
\end{table*}

\section{Simulations}
\label{sim}

\subsection{The large-scale dark matter simulation}
\label{psim}
To find candidate dark matter halos for later refinement we performed a dark matter only simulation of 
a cosmological volume with a comoving side length of 72$\rm Mpc \, h^{-1}$ including $512^3$ dark matter particles 
with individual masses of $m_p=2\times10^8M_{\odot}h^{-1}$. The box is large enough to provide a 
representative piece of the universe and 
the mass resolution fine enough to allow us to reliably find dark matter halos with $\sim 10^3$ particles 
being more massive than $\sim 10^{11} M_{\odot}h^{-1}$. The initial conditions were created using 
GRAFIC1 and LINGERS \citep{1995astro.ph..6070B}, assuming a $\Lambda CDM$ cosmology with nearly 
scale-invariant initial adiabatic fluctuations. The cosmological parameters are based on the 3-year 
results from WMAP \citep{2007ApJS..170..377S} with $\sigma_8$=0.77, $\Omega_m$=0.26, $\Omega_\Lambda$=0.74, 
$h=0.72$ ($\equiv H_{0}$=100$h$ kms$^{-1}$Mpc$^{-1}$) and the initial slope of the power spectrum is $n_s$=0.95.
The initial conditions were then evolved from a redshift of $z \sim 43$ to $z=0$ 
using GADGET-2 \citep{2005MNRAS.364.1105S} with a fixed comoving gravitational softening length of 
$2.52\rm kpc \, h^{-1}$. The simulation data was stored in 95 snapshots separated by $\Delta a=0.01$ 
beginning at a cosmological expansion factor of a=0.06 ($z \approx 43$). 

\begin{figure}
\centering 
\includegraphics[width=7.5cm]{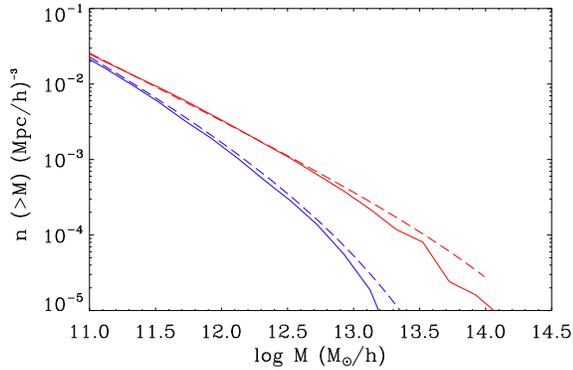}
\caption{Dark matter mass Function (solid) of the $(72h^{-1} Mpc)^3$ box  at z=0 (red) and z=2 (blue). The dashed lines show the 
prediction of \citet{2001MNRAS.323....1S}.}
\label{fig:cmf}
\end{figure}

\begin{figure}
\centering 
\includegraphics[width=7.5cm]{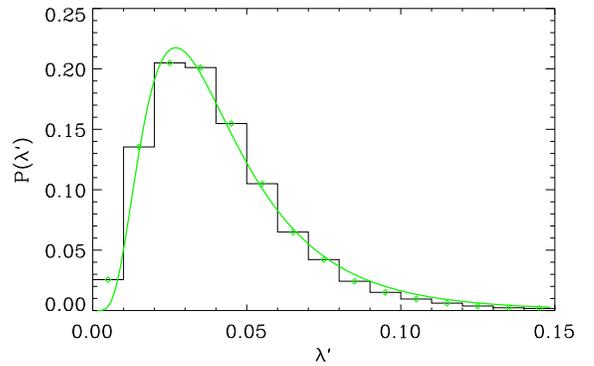}
\caption{Spin parameter distribution for the dark matter box. The green line shows the log-normal-fit with best fit values $\lambda_0'=0.038$ and $\sigma=0.58$.}
\label{fig:lmd}
\end{figure}

At $z$=0 we identify halos with a friends-of-friends algorithm and determine their centers using the 
shrinking sphere technique \citep{2003MNRAS.338...14P}. We then use the radius where the 
mean density drops below 200 times the critical density of the universe ($r_{vir} \equiv r_{200}$) to measure 
the halo mass therein ($m_{vir} \equiv m_{200}$). This results in a complete halo 
catalogue ($n_{\rm halos}=41313$) for halos more massive than $2\times10^{10}M_{\odot}h^{-1}$ which have properties 
typical for this kind of simulation (see \citealp{2010ApJ...710..903M} for a detailed analysis of this simulation). 
In brief, we show the dark matter halo mass function at z=0 and z=2  in Fig. \ref{fig:cmf} along with the 
analytical prediction from \citet{2001MNRAS.323....1S} where we find small variations at the high mass end due
to the limited boxsize. The corresponding distribution of the 
dimensionless spin parameter
\begin{align}
\lambda' \equiv \frac{J}{\sqrt{2}m_{vir}V_cr_{vir}},
\end{align}
defined by \citet{2001ApJ...555..240B}, is shown in Fig. \ref{fig:lmd}. 
Here J is the total angular momentum within $r_{vir}$ and $V_c$ is the halo circular 
velocity $V_c^2=Gm_{vir}/r_{vir}$. The distribution of angular momenta is consistent with previous 
simulations \citep{2001ApJ...555..240B,2002ApJ...581..799V} and can be fitted with a log-normal distribution 
\begin{align}
P(\lambda')=\frac{0.01}{\lambda'\sqrt{2\pi}\sigma} \exp\left(- \frac{\ln^2(\lambda'/\lambda_0')}{2\sigma^2} \right)
\label{eq:lognormal}
\end{align}
with best-fit values $\lambda_0'=0.038$ and $\sigma=0.58$.

\subsection{Refined Simulations}
\label{refsim}
For the higher resolution re-simulations of individual halos we trace all dark matter particles 
that are closer than $2 \times r_{200}$ to the center of the halo at z=0. Following the halo back in 
time we include all particles in the tracing process which are within $2 \times r_{200}$ of the halo 
center at any given snapshot. This ensures that halo encounters during the assembly 
of the halo of interest are always resolved. We found this to be an 
efficient mechanism to reduce contamination with massive boundary particles. The traced particles define the 
region for which we have to generate higher resolution initial conditions. For the cuboid enclosing this region we 
compute the short wavelength modes of the perturbation spectrum using GRAFIC2 \citep{2001ApJS..137....1B}. Based on
the new spectrum we replace the low resolution dark matter particles with dark matter as well as gas particles at 
higher resolution ($\Omega_b$=0.044, $\Omega_{dm}$=0.216). We only consider coherent regions within the cuboid 
that actually contain traced particles. Other regions as well as a 'safety margin' of 1$\rm Mpc \, h^{-1}$ around the high-resolution cuboid are populated 
with particles from the original initial conditions.  To approximate the long range tidal forces, particles 
from the original simulation being further away from the center are merged, with the particle masses increasing as 
the square of the distance from the region of interest. By this and the inclusion of periodic boundaries tidal forces 
from distant regions are accurately included in the computations. 

\begin{figure}
\centering 
\includegraphics[width=7cm]{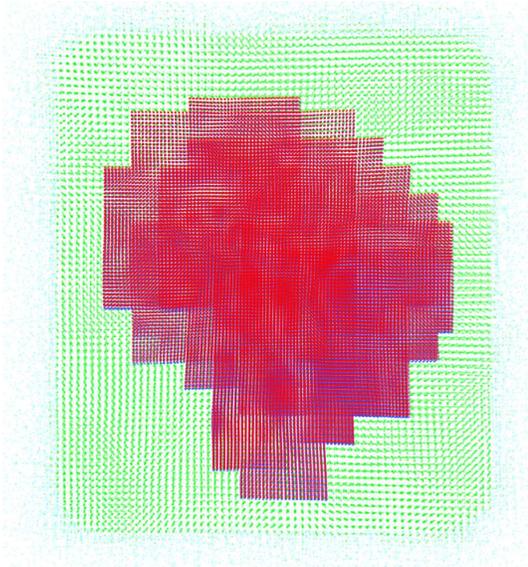}
\caption[Initial Conditions]{Central region of the initial conditions for halo \#0408 at z=43. 
The innermost region consists of the high-resolution gas and dark matter particles 
(red and blue). The green particles are dark matter particles taken from the original 
dark-matter-only run. The outermost dark matter particles have increasing mass depending on 
the distance, with sufficient resolution to represent the long range tidal forces.}
\label{fig:ic}
\end{figure}

We obtain amoeba shaped initial conditions (see Fig. \ref{fig:ic})  for which, on average, approximately 30\% of the high resolution dark matter particles end up inside the 
virial radius at redshift $z=0$ (see \citealp{2003MNRAS.338...14P}  and in particular \citealp{2010MNRAS.403.1859J} 
for alternative ways of creating high resolution initial conditions).
The particle number in the boundary region is kept low enough to perform the simulations 
in a reasonable time. For example the most massive halo \#0040 which has a total mass $m_{200}$ of $2.7 \times 10^{13} 
M_{\odot}h^{-1}$ at $z=0$ took $\sim 23000$ CPU-hours to evolve ($3.8 \times 10^{6}$ high-resolution particles in 
dark matter and gas each). In the re-simulations the particles in the high resolution regions have a 
gas and star mass of $m_{*,gas}=4.2 \times 10^{6}M_{\odot}h^{-1}$ (we spawn one star per gas particle) and
a dark matter mass of $m_{dm} = 2.5 \times 10^{7}M_{\odot}h^{-1}$ which is 8 smaller than in the original simulation.
The comoving gravitational softening length for the gas and star particles is $400 \rm pc \, h^{-1}$ and $890 \rm pc \, h^{-1}$
for the high resolution dark matter particles, scaled with the square root of the mass ratio \citep{2001MNRAS.324..273D}.
Compared to some other recent cosmological zoom simulations 
\citep{2009MNRAS.396..696S,2009MNRAS.398..312G,2009arXiv0909.4167P, 2010ApJ...709..218F} the resolution level 
of our simulations at $M_{halo} \approx 10^{12}M_{\odot}$ is slightly lower. But while these simulations are limited to a few halos 
in a small mass range we performed a significantly larger number of re-simulations of halos spanning a mass 
range of almost two orders of magnitude. The present-day properties of our re-simulated galaxies can be found in Table \ref{tab:ins-acc}.
Finally, we also performed a number of re-simulations at higher resolution, i.e. with particle masses 8 times lower and half the softening length. While increasing resolution can slightly change the individual accretion histories of the galaxies, the global trends found
in this paper remain the same. 
%In agreement with the results of \citep{2007ApJ...658..710N} we find that late time in-situ star formation is reduced at higher resolution.

%The addition of small scale power in the fluctuation spectrum as well as the introduction of gas 
%physics naturally leads to somewhat different final dark matter properties at z=0. In some sense, we run a 
%new simulation that should be similar but not identical to the original. 
%However the changes in  $r_{200}$ and  $m_{200}$ are small, but sometimes we observe
%differences in the merger history of a halo, e.g. the time at which the mergers occur is shifted. 
%Most mentionable, halo 3431 has a large companion (M $\sim 10^{11} M_{\odot}/h$) in the re-simulations 
%at z=0, whereas in the original run those two halos have already merged by z=0. 
%{\bf this could also be an effect of our initial conditions}
%For completeness we performed all of our re-simulations without gas particles.
%The virial properties of the 39 re-simulated halos with and without gas are listed in table \ref{tab:haloprop}. 

\subsection{Simulation details}
\label{codepar}
%Star Formation prescription and feedback Springel&Hernquist 2003
The simulations presented here have been performed using the parallel TreeSPH code GADGET-2 
\citep{2005MNRAS.364.1105S}
which calculates the gas dynamics using the Lagrangian Smoothed Particle Hydrodynamics technique (SPH, see e.g.
\citealp{1992ARA&A..30..543M}).
The code ensures the conservation of energy and entropy \citep{2002MNRAS.333..649S} and includes star formation 
and cooling for a primordial composition of hydrogen and helium, where the cooling rates are computed
under the assumption that the gas is optically thin and in ionization equilibrium \citep{1996ApJS..105...19K}. 
Additionally, the simulations include a spatially uniform redshift-dependent UV background radiation field 
with a modified \citealp{1996ApJ...461...20H} spectrum, where reionization takes place at 
$z \approx 6$ \citep{1999ApJ...511..521D} and the UV background radiation field peaks at $z \approx 2-3$.
For a recent detailed investigation on the effects of varying the background radiation field on the evolution of galaxies,
see e.g. \citep{2010arXiv1009.6005H}.

For the star formation and feedback prescription we use the self-regulated supernova feedback 
model of \citealp{2003MNRAS.339..289S}. This models treats the ISM as a two-phase medium 
\citep{1977ApJ...218..148M, 2006MNRAS.371.1519J} where clouds of cold gas are embedded in the hot gas
 phase at pressure equilibrium. Stars are allowed to form out of the cold gas phase if the local 
density exceeds a threshold value ($n > n_{\rm th} = 0.205\rm cm^{-3}$) which is calculated self-consistently 
in a way that the equation of state is continuous at the onset of star formation. In this model the 
star formation rate is set by $d\rho_*/dt = (1 - \beta)\rho_c/t_*$, where $\beta$ is the mass fraction of 
massive stars ($>8M_{\odot}$), $\rho_c$ is the density of cold gas and $t_*$ is the star formation time scale set by
$t_*=t^0_*(n/n_{th})^{-1/2}$. The massive short-lived stars heat up the surrounding gas with an energy input 
of $10^{51}ergs$ due to supernovae. In order to prevent spurious star formation at high redshift we 
require an over-density of $\Delta > 55.7$ for star formation to set in. 

\subsection{The baryonic mass budget}
\label{barfrac}
\begin{figure}
\centering 
\includegraphics[width=8cm]{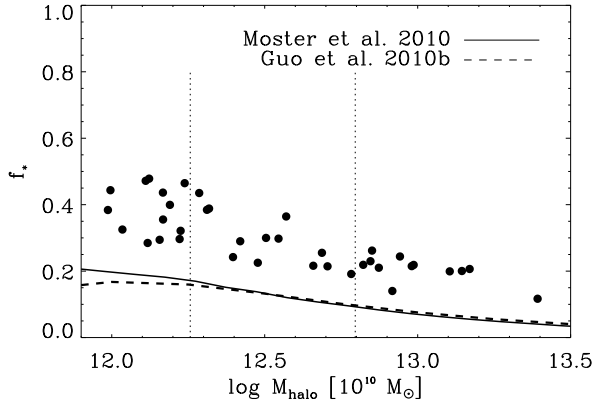}

\caption[]{Fraction of baryons that is converted into stars at redshift zero. The vertical dotted lines indicate 
the separation into the different mass bins. The solid black line shows the results of \citet{2010MNRAS.404.1111G}, 
the dashed line those of \citet{2010ApJ...710..903M}.
% and the dotted line those of \citet{2006ApJ...643...14S}.
} 
\label{fig:barfrac}
\end{figure}

Using similar parameters for zoom simulations has been shown to result in galaxies with reasonable present day 
properties \citep[][2010 in prep.]{2007ApJ...658..710N,2009ApJ...699L.178N,2009ApJ...697L..38J}.
However, the employed star formation prescription favors efficient star formation at early times resulting in preferentially spheroidal
systems with old stellar populations, due to the strongly self-regulated feedback which does not produce the supernova driven winds.
Fig. \ref{fig:barfrac} shows the conversion efficiency of the simulated galaxies at the present day 
$f_* = m_*/(f_b*m_{vir,dark})$ where $m_*$ is the stellar mass 
within 10 \% of the virial radius, $f_b=\Omega_b/\Omega_{dm} =0.20$ is the cosmic baryon fraction and $m_{vir,dark}$ is the dark matter mass within the 
virial radius of the galaxy. Therefore $f_b * m_{vir,dark}$ is the amount of total baryonic matter 
available in each halo and $f_*$ the fraction thereof that is converted into stars in the central galaxy.
This fraction declines in a roughly linear fashion with the logarithm of the 
halo mass from $f_* \approx 0.5$ for the smallest halos ($\approx 10^{11.9} M_{\odot}$) to $f_* \approx 0.15$ for high 
mass halos ($\gtrsim 10^{13} M_{\odot}$) , still over-predicting by a factor of 2
%(factor of 5-8 for low mass halos and 2-3 for high mass halos) 
the estimation from recent models (see however \citet{2006ApJ...643...14S} who find higher efficiencies for high mass
galaxies) that are tested by matching observed luminosity functions to 
simulated halo mass functions \citep{2010ApJ...710..903M,2010MNRAS.404.1111G,2009ApJ...696..620C,2010ApJ...717..379B} or weak lensing observations
\citep{2006MNRAS.368..715M}. Note that a Salpeter initial mass function would increase the 'observed' conversion efficiency
by approximately a factor of two \citep{2010arXiv1009.5992V}.

The physical processes probably responsible for this discrepancy are well studied and it has been argued that feedback from 
SNII is important for low mass systems \citep[e.g.][]{1974MNRAS.169..229L,1986ApJ...303...39D,2010arXiv1006.0106G} and 
feedback from 
supermassive black holes dominates for high mass systems \citep{2006MNRAS.365...11C,2008ApJ...676...33D}.
Although this issue is relatively well understood and 
many idealized calculations have shown how these feedback processes can expel the baryons 
from galaxies, there have been only a few high resolution galaxy formation calculations, using 
cosmological initial conditions, beginning to master the physics well enough to match either the 
winds seen in forming galaxies or the final metal distribution between galaxies and the IGM 
\citep{2008MNRAS.389.1137S,2010MNRAS.402.1599S}. Some other calculations do successfully allow for 
winds and for the consequences these winds have on the galaxies and the surrounding ISM  
\citep{2003MNRAS.339..289S,2008MNRAS.387..577O,2010MNRAS.tmp..860O,2010arXiv1005.1451C,
2010arXiv1005.3921W,2010MNRAS.tmp..740M}. Our computations do not generate significant winds at high redshift 
(e.g. \citealp{2010ApJ...717..289S}) and thus 
overestimate, by roughly a factor of two, the condensed baryon fraction of massive 
galaxies \citep{2010MNRAS.404.1111G,2010ApJ...710..903M}. 
This becomes worse if we extend the sample to lower masses where the halo occupation models
predict a sharp drop off the conversion efficiency $f_*$. This is probably due to the fact that ejective supernovae wind feedback, which
is not included in the present study, is most effective in this regime.
We are currently working to implement physically valid feedback implementations to address this problem.

\begin{figure}
\centering 
\includegraphics[width=8cm]{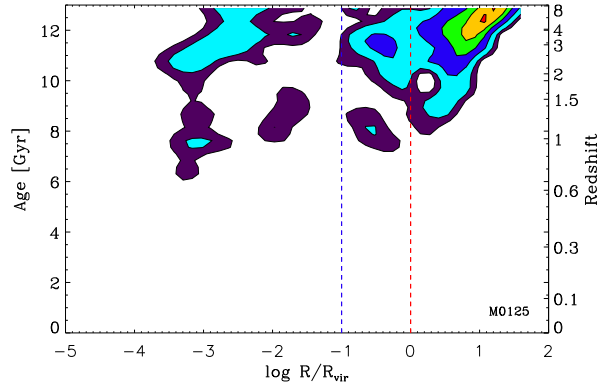}

\caption[]{Stellar origin diagram for all stars within $r_{10}$ at $z=0$ in galaxy M0125. Every grey dot indicates
the time when a stellar particle was born and the distance, in units of the virial radius of the main galaxy 
at that time, where it was born. The contours enclose 90\% (purple), 80\%(turquoise), 60\% (blue), 40\% (green), 
25\% (orange) and 10\% (red) of the stars, respectively. 
The blue and red vertical lines show the virial radius and 10\% of the virial radius, 
respectively. There is a clear distinction between stars initially formed in the galaxy and those 
formed outside the galaxy and are accreted later on (78 percent of all stars).\\
(Dots excluded for file size reasons)} 
\label{M0125.ins-age}
\end{figure}

\begin{figure}
\centering 
\includegraphics[width=8cm]{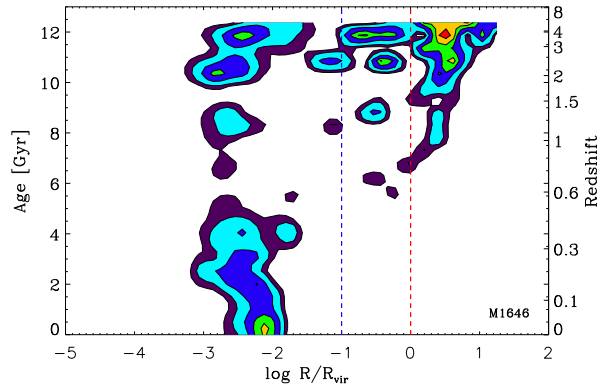}
\caption[]{Same as Fig. \ref{M0125.ins-age} but for the low mass galaxy M1646. There is significant 
in-situ star formation at the center even at low redshift and significantly less accretion of stars. 
In this case only 52 per cent of the stars are accreted.\\(Dots excluded for file size reasons)} 
\label{M1646.ins-age}
\end{figure}

\begin{figure}
\centering 
\includegraphics[width=8.0cm]{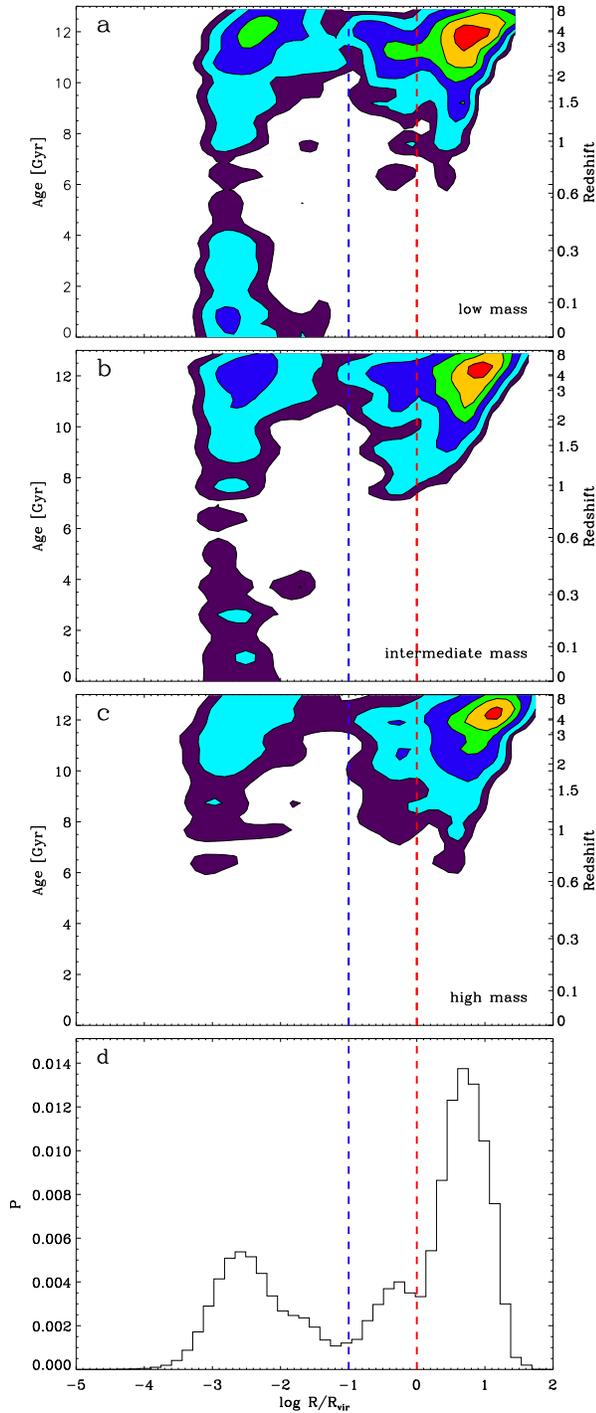}
\caption[]{Same as Fig. \ref{M0125.ins-age} but for all galaxies in low mass halos in the mass 
range $7.0 \times 10^{11} - 1.3 \times 10^{12}h^{-1} M_{\odot}$ (panel a), for intermediate halo masses 
in the range $1.3 \times 10^{12} - 4.5 \times 10^{12} h^{-1} M_{\odot}$ (panel b), and for all high mass 
halos with $4.5 \times 10^{12} - 2.7 \times 10^{13} h^{-1} M_{\odot}$ (panel c). 
%The contours enclose 90\% (purple), 80\%(turquoise), 60\% (blue), 40\% (green), 25\% (orange) and 10\% (red) of the stars, respectively. 
The contours show the same percentiles as in Figs. \ref{M0125.ins-age} and \ref{M1646.ins-age}.
The stars form in two phases, either inside $r_{10}$ or outside $r_{\mathrm{vir}}$ as can 
be seen in panel d. Galaxies in low mass halos have ongoing in-situ star formation (see Fig. \ref{fig:sfr}) 
at relatively high specific rates until the present day, whereas in the highest mass group most star 
formation is complete by z=2.}
\label{in-situ_mult}
\end{figure}

\section{The two phases of galaxy formation} 
\label{twophases}

\begin{figure}
\centering 
\includegraphics[width=8.5cm]{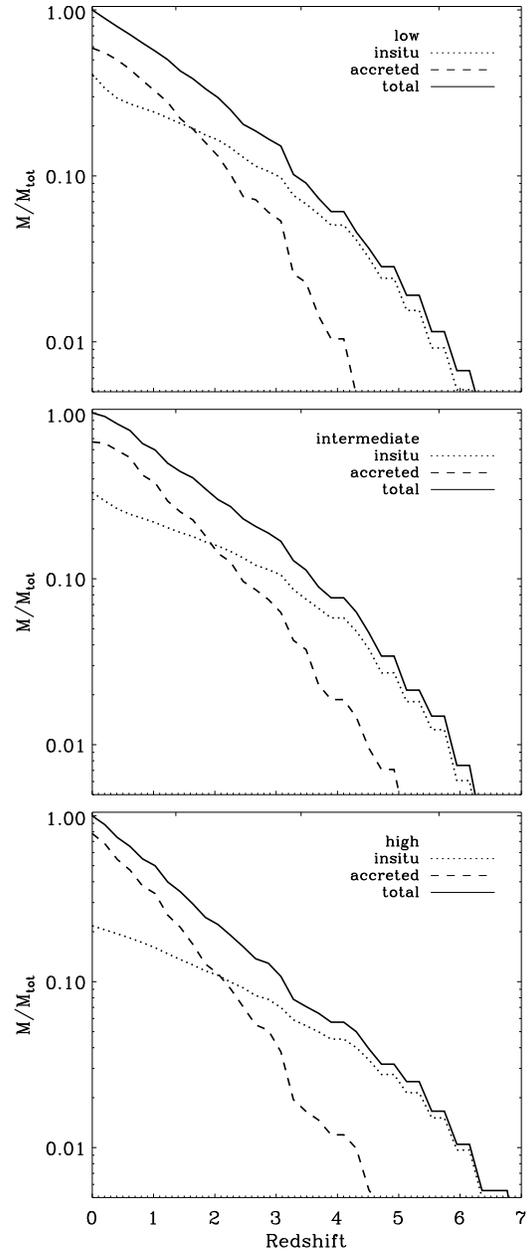}
\caption[]{Stellar mass assembly histories (solid lines) for low mass (top), intermediate mass 
(middle) and high mass (bottom) galaxies. The assembly is separated into in-situ stars 
(dotted line) and ex-situ stars that are accreted onto the galaxy later on (dashed line). The assembly of
higher mass galaxies is more dominated by in-situ formation at high redshift, however, the total 
fraction of accreted stars by z=0 is higher ($\approx 80\%$) for massive systems than for low mass 
systems ($\approx 60\%$).}
%{\bf -- error bars}}
\label{acchist_mult}
\end{figure}

\begin{figure}

\centering 
\includegraphics[width=8.0cm]{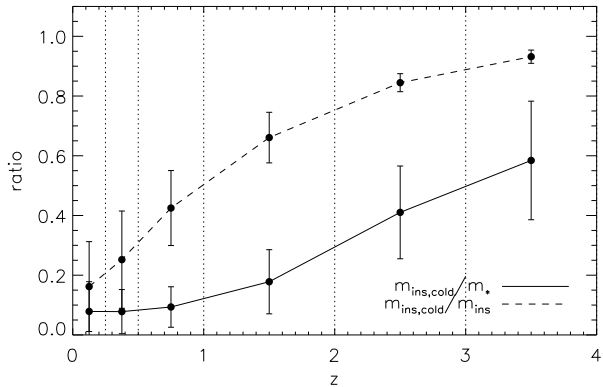}
\caption{Average ratio of in-situ created stars that formed inside the bins indicated by the vertical
dotted lines out of gas that was accreted cold to the total 
mass of in-situ created stars (dashed line). The solid line shows the ratio
of the stars created in-situ out of cold gas to the total stellar mass growth, 
this includes in-situ star formation as well as accretion. The error bars corespond to the
1$\sigma$-dispersion.}
\label{fig:coldfraction}

\end{figure}

\begin{figure}

\centering 
\includegraphics[width=8.5cm]{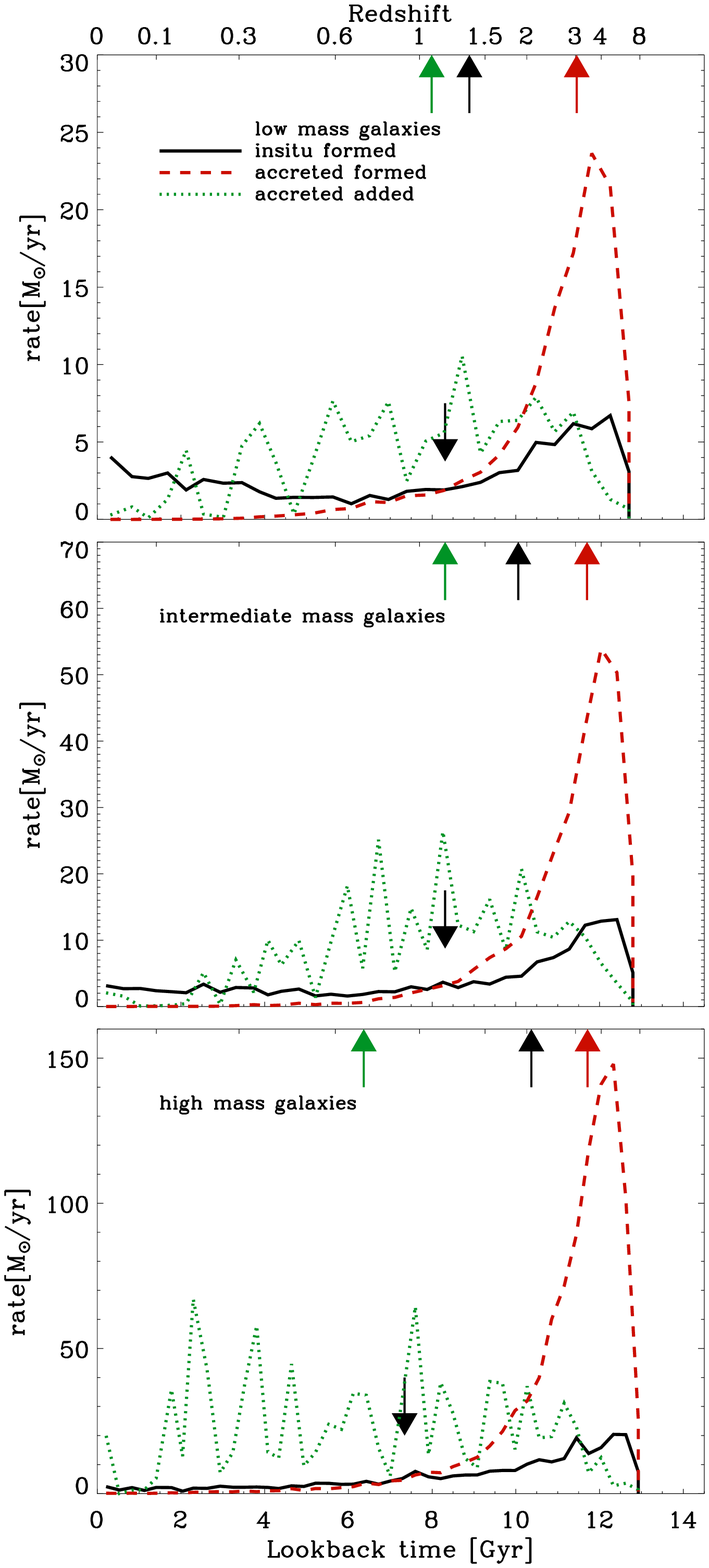}
\caption{Star formation histories for low mass (upper panel), intermediate mass (middle panel) and 
high mass galaxies (lower panel) for all stars that end up inside the galaxy at $z=0$.
The solid black line shows the formation of the in-situ created stars, the red dashed line the formation of 
the ex-situ stars and the green dotted line shows the accretion rate of the ex-situ stars onto the galaxy.
The arrows on top indicate the time at which half the stars are formed/added.
The arrow at the bottom indicates the time at which 50\% of the final galaxy mass is assembled.} 
\label{fig:sfr}

\end{figure}

The stellar particles ending up in the simulated galaxies at $z=0$ are of two 
different origins. Some fraction of the stars are made in-situ, within the galaxies, from accreted 
gas and some fraction of the stars are made ex-situ outside the galaxies and are accreted later on 
\citep{2007ApJ...658..710N,2009ApJ...697L..38J}. The relative amount of in-situ and ex-situ stars 
is found to vary systematically with galaxy mass. Two typical stellar origin diagrams indicating this behavior are 
shown in Fig. \ref{M0125.ins-age} and Fig. \ref{M1646.ins-age}.

\begin{figure}
\centering 
\includegraphics[width=7.5cm]{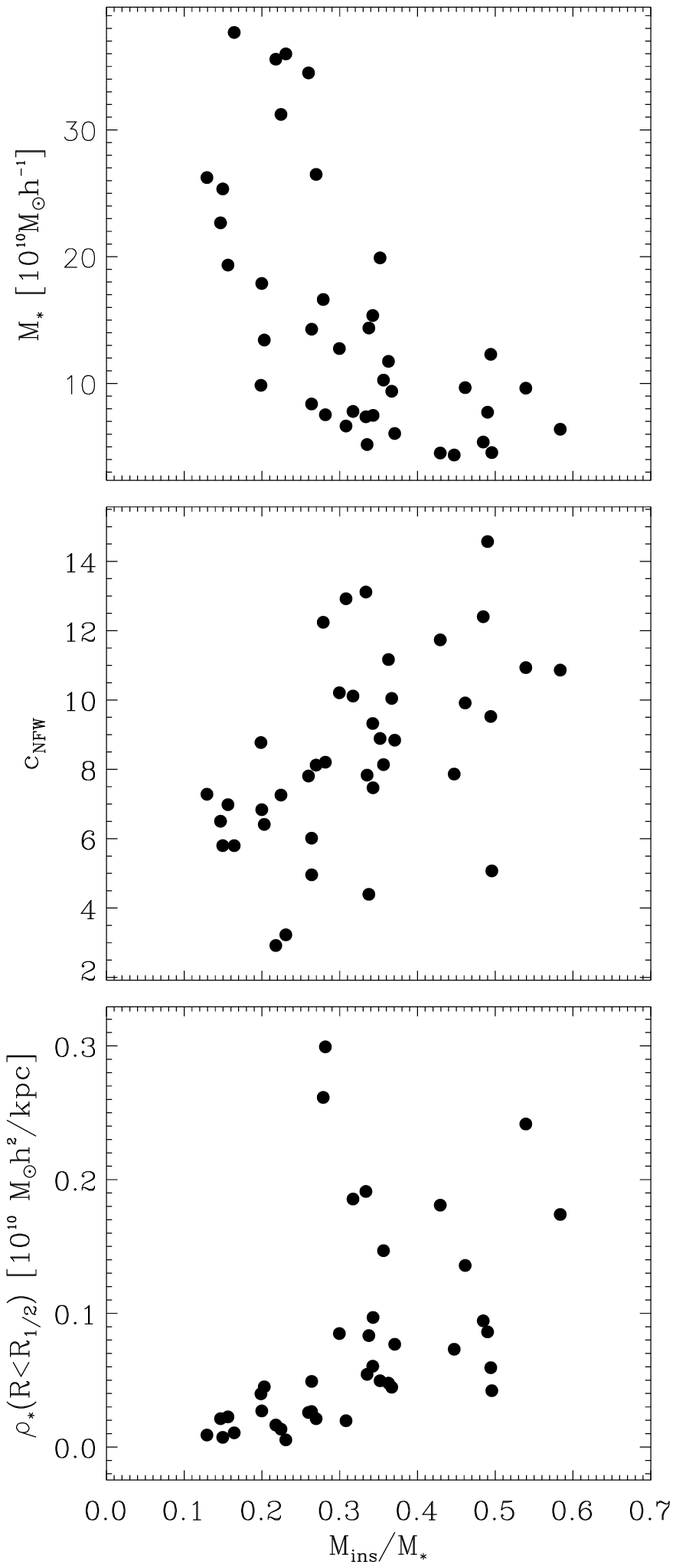}
\caption[]{From top to bottom: Fraction of in-situ stellar mass vs total stellar mass inside $r_{10}$, 
halo concentration and stellar density inside $R_{1/2}$ at $z=0$. There is a clear trend that galaxies with less
in-situ star formation are more massive, have less concentrated halos and lower density central regions.}
\label{gmass-c-od} 
\end{figure}

To construct these diagrams we follow every star that ends up within 10\% of the present-day virial radius of a 
simulated galaxy back in time. We use 10\% of the virial radius, $r_{10}$, as a fiducial value for the extent 
of the stellar component of a simulated final galaxy inside its dark matter halo. Then we mark the time when a star 
was born as well as its distance from the galaxy center in units of the virial radius (at this time) with a grey dot. 
The values are discrete in time representing the discrete snapshots. The contours in these plots encompass the smallest
number of bins that include 10, 25, 40, 60, 80 and 90 per cent of the stars, respectively. In Fig.\ref{M0125.ins-age} we 
show the stellar origin diagram for a massive system with a halo mass of $\sim 10^{13}M_{\odot} h^{-1}$. At redshifts $z>2$ 
there are two separate peaks of star fomation: one inside $r_{10}$, which is in-situ star formation and another one outside 
the virial radius of the system at that time. This indicates that a significant fraction 
of the stars in the present-day galaxy was made outside the galaxy and has been accreted later on. 
For this system the in-situ star formation decreases towards lower redshifts. Although there is ongoing star formation
until z=0 the contribution to the final galaxy is negligible, since the contoured regions include 90 per cent 
of all stars in the galaxy. For a lower mass system 
with a halo mass of $7.1\times10^{11} M_{\odot} h^{-1}$ the same analysis is shown in Fig. \ref{M1646.ins-age}.
In this case the fraction of stars forming ex-situ is lower and the contoured regions extend up to the present day, i.e.
in-situ star formation continues at a significant level towards lower redshift. 

In Fig. \ref{in-situ_mult}, we have stacked all simulated galaxies of our sample into three mass bins 
(indicated by the horizontal bars in table \ref{tab:ins-acc})
% and \ref{tab:haloprop}) 
with the same number of objects (13), every star particle is weighted according to the total number of stars in its host
galaxy, so that every galaxy has an equal weight. The low mass bin contains galaxies with halo masses in the range 
$7.0 \times 10^{11} - 1.3 \times 10^{12} M_{\odot} h^{-1}$ (panel a), 
intermediate mass galaxies have $1.3 \times 10^{12} - 4.5 \times 10^{12} M_{\odot} h^{-1}$ (panel b) and high mass galaxies 
have $4.5 \times 10^{12} - 2.7 \times 10^{13} M_{\odot} h^{-1}$ (panel c). These plots again demonstrate in a 
more statistical sense that the stars ending up in the final galaxies form in two distinct phases, 
namely in-situ in the galaxy and ex-situ outside the virial radii of the galaxies (red vertical dashed lines).
The spatial division line between these two phases of star formation is at about 10\% 
of the virial radius indicated by the vertical blue dashed lines in Fig. \ref{in-situ_mult}.
In addition, there is a clear trend that low mass galaxies have relatively more in-situ star formation at 
low redshift $z<1$ than higher mass galaxies. For the most massive galaxies the contribution from 
late in-situ star formation is relatively small. 
Panel d shows a histogram for the formation radii for all stars in all simulations. 
For this analysis we use 45 logaritmically evenly spaced bins. We see two peaks, for the in-situ created stars at 
$\log ( \rm r/r_{vir}) \approx -2.5$ and for the ex-situ created stars at $\log ( \rm r/r_{vir}) \approx 0.6$, respectively.
A third peak appears between $\rm r_{10}$ and $\rm r_{vir}$ that is due to infalling substructure that is still star-forming.

In Fig. \ref{acchist_mult} we show the average mass accretion histories for the stellar particles in 
the three mass bins separated into in-situ and ex-situ/accreted stars depending on whether they have formed
inside or outside 10\% of the virial radius. The galaxy growth  is dominated for all three mass bins by 
in-situ star formation until $z \approx 2$, when the mass of accreted stars equals the mass of in-situ stars. By $z=0$
about $41 \pm 9\%$ (we give mean values and the 1$\sigma$-dispersion of the 13 galaxies) of the stars in the low mass sample 
(top panel) have formed in-situ, 
the rest were accreted. For the  intermediate mass galaxies (middle panel) 
%the growth is dominated by accretion since $z\approx 2$. The present-day {\bf intermediate mass} galaxies have a lower 
the fraction of in-situ stars is lower than for the low mass sample of $\approx$ $33 \pm 10\%$, and $67 \%$ of 
the stars were accreted. With $78 \pm 7 \%$ the fraction of accreted stars is even higher for the massive galaxies. 
On average only $22 \%$ of the present-day stellar mass is formed in-situ which is the dominant mode until $z \approx 2$ 
but thereafter contributes very little to the stellar mass growth.

Following \citet{2005MNRAS.363....2K} and \citet{2009MNRAS.396.2332K} we examined whether the gas out of which the in-situ
stars are formed in our galaxies was ever heated above $T_{hot}>2.5\times 10^5K$ throughout the simulation.
The results can be seen in Fig. \ref{fig:coldfraction}. The dashed line shows, that up to redshift 2, where in-situ
star formation is still dominating over accretion, almost all of the in-situ stars are formed out of gas that was 
accreted cold. Only at later times ($0<z<2$), when stellar accretion is the primary source of stellar mass growth, in-situ
stars are forming out of cooling hot halo gas. At lower redshift the contribution of in-situ star formation out of cold flows
to the total stellar mass growth becomes almost negligible (dotted line in Fig. \ref{fig:coldfraction}). 
The interpretation of the results does not change when we instead of a fixed temperature cut use a temperature threshold 
related to the current halo virial temperature (see \citet{2005MNRAS.363....2K}).
This is in agreement
with the previous results of numerical simulations \citep{2009MNRAS.396.2332K} and analytical predictions \citep{2006MNRAS.368....2D}
that galaxy growth at high redshift ($z \geq 2  $) is dominated by cold accretion.

Fig. \ref{fig:sfr} illustrates the star formation and assembly histories for the galaxies in the 
three mass bins. The red dashed line shows the archaeological star formation history of the accreted stars 
computed from the mass weighted 
ages of the accreted stars at the present day. All curves show a steep increase towards the peak 
at $z\approx 4$ at values of $\approx 25 M_{\odot} yr^{-1}$, $\approx 55 M_{\odot} yr^{-1}$, and 
$\approx 150 M_{\odot} yr^{-1}$ for the low, intermediate and massive bin, respectively. This is followed by 
an approximately exponential decline towards $z=0$.  
The red arrow on top indcates the time when half of the accreted stars are formed. In all 
cases, i.e. at all masses this is at $z\approx3$. The green dotted line shows when these stars 
are accreted onto the galaxies. As this happens in mergers, the curves show peaks. On average the rates 
increase towards $z=2$ and then stay relatively flat with average rates of 
$\approx 3.6 M_{\odot} yr^{-1}$, $\approx 8.2 M_{\odot} yr^{-1}$, and $\approx 17 M_{\odot} yr^{-1}$. 
The green arrow on top indicates when half 
of the present-day mass in ex-situ stars is accreted onto the galaxies. This happens around z=0.7-1.2 
and therefore significantly later than the formation of these stars at z=3-4. 
The black solid line shows the formation history of the in-situ stars in the galaxies. This is most 
closely related to the star formation rate that would actually be observed in these galaxies. All curves peak 
at $ z \ge 3.5$ at rates between $\approx 5$ and $\approx 20 M_{\odot} yr^{-1}$. Independent of galaxy mass
all rates drop to $\approx 2-3 M_{\odot} yr^{-1}$ at $z=1$ and stay constant to the present day similarly to the observations
of massive galaxies by \citet{2005ApJ...619L.135J}. 
This results in a specific star formation rate of $0.31 \pm 0.15$, 
$0.18 \pm 0.15$ and $0.053 \pm 0.071 \times 10^{-10}yr^{-1}$ for the different mass bins. 
According to the definition by \citealp{2008ApJ...688..770F} ($SFR/m_* < 0.3 / t_{thub}$) the galaxies in the high mass
bin would correspond to quiescent galaxies.
The time when half of the in-situ 
stars are formed is indicated by the top black arrows. This changes systematically with galaxy mass from z=1.4 to 
z=1.9 and z=2.1, i.e. the in-situ component is oldest for the most massive 
systems.  The black arrow at the bottom of the panels indicates the time when half of the final galaxy was 
assembled. For all galaxies this is around redshift $z \approx 1$. Therefore all galaxies double their mass 
since then. For low mass systems the low redshift growth is dominated by in-situ formation whereas for high mass 
systems it is dominated by accretion of small stellar systems \citep{2010arXiv1009.5185T}. 

\begin{figure}
\centering 
\includegraphics[width=7.5cm]{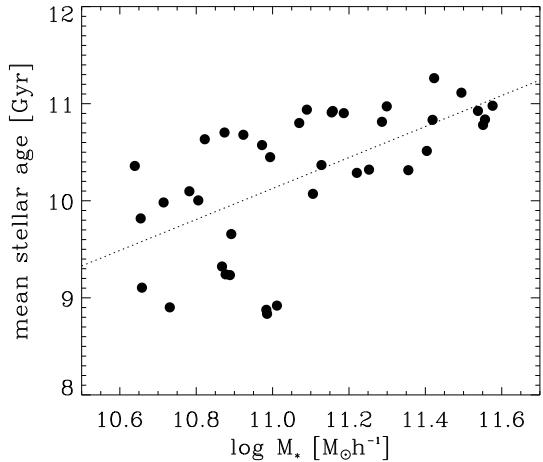}
\caption[]{Mean age of the stars inside $r_{10}$ as function of galaxy mass. High mass galaxies 
consist of older stars than the low mass galaxies, recovering the phenomenon usually 
referred to as 'archaeological downsizing' ($t_{mean} \propto \log \, M_*^{1.6} $).}
\label{fig:gmass-age} 
\end{figure}

%\begin{figure}
%\centering 
%\includegraphics[width=7.5cm]{./gm-z50.eps}
%\caption[]{Redshift when the galaxies have assembled 50\% of their present-day stellar mass as function of galaxy ma%ss.
%Although the stars in the massive galaxies are typically older (see Fig. \ref{fig:gmass-age}) there is a trend that %the massive galaxies
%are assembled later. This is consistent with the global picture of hierarchical structure formation. 
%}
%\label{fig:gmass-z50} 
%\end{figure}

In summary, at high redshift the assembly of galaxies at all masses is dominated by in-situ star formation fed by cold flows. 
The larger the galaxy mass
%the earlier accretion starts to dominate the stellar mass assembly and 
the smaller is the late contribution of in-situ star formation. 
At low redshift, $z <1$, the growth of low mass galaxies continues by in-situ star formation and stellar accretion 
whereas, massive galaxies grow predominantly by accretion of ex-situ stars (see e.g. \citealp{2010ApJ...709..218F,2009ApJ...699L.178N}). 

In Fig. \ref{gmass-c-od} we show interesting correlations of galaxy and halo properties with the fraction of 
in-situ stars indicating that this quantity is an important tracer of galaxy assembly.
Essentially, this ratio,  $m_{ins}/m_*$, is a dimensionless measure for the degree to which the galaxy was formed 
by a dissipational versus a dissipationless process \citep{2010ApJ...712...88L}.
The fraction of the stellar galaxy mass formed in-situ $m_{ins}/m_*$ is highest, up to 60\%, for low mass galaxies
and declines almost linearly (despite some scatter) with increasing galaxy mass down to $\approx$ 13\% for 
the most massive systems in our simulations which are the central galaxies of massive groups 
(top panel of Fig. \ref{gmass-c-od}). This trend is very similar to semi-analytical predictions \citep{2006ApJ...648L..21K}
and constraints based on halo occupation models combined with isolated merger
simulations \citep{2009ApJ...691.1424H}.
In the central panel of Fig. \ref{gmass-c-od} we show the fraction of 
in-situ mass versus the concentration parameter $c$ of the dark halo which is defined as the ratio between 
$r_{200}$ and $r_{s}$, where $r_s$ is the scale radius for an NFW fit \citep{1997ApJ...490..493N} of the density 
profile:
\begin{align}
\rho(r) = \frac{\delta_c \rho_{crit}}{(r/r_s)(1+r/r_s)^2}
\end{align}
For the fit we binned the halo into 32 spherical shells equally spaced in $log_{10}(r)$ 
between $r_{200}$ and $log_{10}(r/r_{200})=-2.5$ similar to \citet{2009MNRAS.394.1559G}. We see a continuous change 
of the dark matter halo concentration. As expected from the effect of adiabatic contraction  galaxies with 
significant in-situ star formation, i.e. more dissipation, live in more concentrated halos 
\citep{1986ApJ...301...27B,1994ApJ...431..617D,2002ApJ...571L..89J,2008ApJ...681.1076D,2004ApJ...616...16G,2010MNRAS.tmp..847A,2010arXiv1007.2409A}. The concentration of more massive halos does not increase significantly as the matter is added 
predominantly in stellar form and cannot dissipate (see e.g. \citealp{2009ApJ...697L..38J} and references therein),
i.e. the adiabatic contraction approximation cannot be applied for massive galaxies.
The bottom panel in Fig. \ref{gmass-c-od} shows the stellar density inside the spherical half-mass radius versus the 
ratio of in-situ created stars of the galaxies. The two properties are correlated in the sense that 
galaxies with a large fraction of accreted stars have lower central densities, a well known property of 
elliptical galaxies \citep[e.g.][]{1992ApJ...399..462B}.

\begin{figure*}
\centering 
\includegraphics[width=17.5cm]{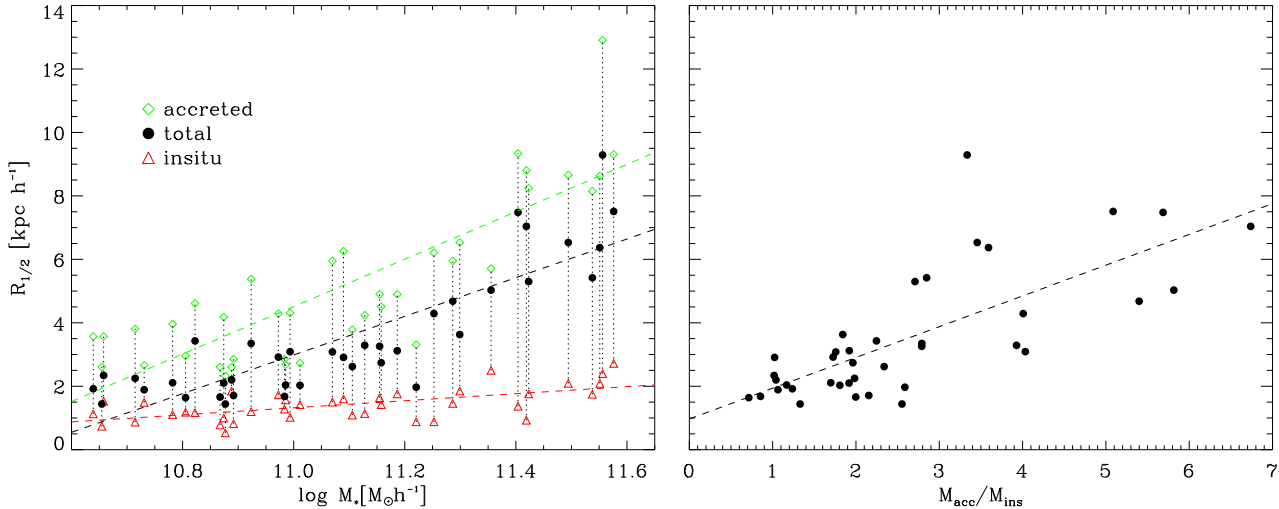}
\caption{{\it Left panel:} 
Stellar mass inside $10\%$ of the virial radius vs. spherical half-mass radius of accreted (green diamonds), in-situ (black triangles) and all stars (red squares), 
respectively. The dashed lines show the results of a linear fit for the respective components   
($r_{1/2} \propto \log \, M_*^{\alpha}$, with $\alpha =$ 7.5, 6.1 and 1.1 for the accreted, total and in-situ stars, respectively).
While the half-mass radius of the accreted stars strongly increases with mass, the half-mass radius 
of the in-situ formed stars shows only a weak dependence on galaxy mass. The mass-size relation is driven by the 
by the accreted stars. 
{\it Right panel:} This plot shows the spherical half-mass radii of the galaxies as a function of the ratio of accreted to 
in-situ created stars. The size increase of the galaxies is roughly linear dependent on this ratio ($r_{1/2}  \propto 0.97 * M_{acc}/M_{ins} $).}
\label{fig:halfrad}
\end{figure*}

Fig. \ref{fig:sfr} gives a clue to the paradox of 'downsizing'. The initial expectation was that in a hierarchical
universe, since more massive halos statistically are formed later than less massive ones, the same should 
be true of galaxies. But we know that this is not true observationally \citep{2005ApJ...632..137N}, giant ellipticals are older - not younger - than lower mass systems (see e.g. \citealp{2005ApJ...621..673T}).
Our simulations give the same result as can be seen from Fig. \ref{fig:gmass-age}, the most massive systems are
made out of the oldest stars. The inclusion of galactic winds would probably lead to less efficient
star formation at high redshifts and leave more gas for late in-situ star formation especially in the lower mass systems, rendering these galaxies
even younger. This would lead to an even steeper relation than the one shown in Fig. \ref{fig:gmass-age}.
The explanation of the paradox is obvious: The accreted stars are 
typically made in smaller systems and these small systems are in fact made at early times (dashed red curves in 
Fig. \ref{fig:sfr}). Massive galaxies are more dominated by the accreted stars and so by z=0 they contain 
primarily old stars, although the galaxies themselves are assembled late. \citealp{2006MNRAS.366..499D} obtain the same result with
their semi-analytic model. 
This way the expectation from hierarchical structure formation is satisfied.
Both our simulations and the observations of \cite{2008ApJ...677L...5V} agree: even at late times massive galaxies continue to grow in mass and size.

\section{Galaxy sizes}
\label{galsizes}

%\begin{figure*}
%\centering 
%\includegraphics[width=17.5cm]{./rad-lin.eps}
%\caption{{\it Left panel:} 
%Stellar mass inside $10\%$ of the virial radius vs. half-mass radius of accreted (green diamonds), in-situ (black triangles) and all stars (red squares), 
%respectively. The dashed lines show the results of a linear fit for the respective components.   While the half-mass radius of the accreted stars strongly increases with mass, the half-mass radius 
%of the in-situ formed stars shows only a weak dependence on galaxy mass. The mass-size relation is driven by the 
%by the accreted stars. 
%{\it Right panel:} This plot shows the half-mass radii of the different components as a function of the fraction of stellar mass that is accreted by the galaxy. {\bf T: This plot looks weird, maybe use a linear scale for the x-axis?} }
%\label{fig:}
%\end{figure*}

The left panel of Fig. \ref{fig:halfrad} shows the present-day spherical half-mass radius for the different 
components of our galaxies. The size of the in-situ component shows a very weak trend with galaxy mass. 
For the low mass galaxies the half-mass radii of the in-situ and the accreted stars are of similar size. 
While the in-situ component does not get larger than $\approx 3 \rm kpc \, h^{-1}$, the half-mass radius of the accreted stars is 
strongly increasing with galaxy mass and since the fraction of accreted stars rises with galaxy mass as well, 
the global half-mass radius of the galaxies follows this trend. 
In our simulations the majority of the in-situ created stars are formed in the bulges of the galaxies. 
Stronger feedback mechanism would probably lead to more star formation in galactic disks resulting in larger radii
of the in-situ component. The half-mass radius of the accreted stars should not be affected by this.
The right panel of Fig. \ref{fig:halfrad} shows the galaxy radii versus the ratio of
accreted to in-situ created stars. We find an almost linear trend. 
Fig. \ref{fig:halfrad} shows, that stellar accretion is the dominant mechanism for the size growth of massive galaxies.
For most of our systems the accretion of stars is significant at low redshifts, as seen in Fig. \ref{fig:sfr}, 
especially for the high mass galaxies. Half of the total accreted stellar mass is added to the galaxies 
between redshift one and the present day which leads to considerable size increase at late times.
Consistent with this pictures are the observations from e.g. \cite{2008ApJ...677L...5V} and others
that show a significant growth between redshift $z=3$ and $z=0$ for quiescent early type galaxies.
A detailed analysis of this effect will be presented separately. We give, in \cite{2009ApJ...699L.178N} 
a simple argument based on the virial theorem showing how late accretion of low mass satellites
('minor mergers') will lead to the rapid growth in galactic size.
%However, the detailed study of the size evolution of our galaxies will be subject to a following paper.

\section{Summary and Discussion}
\label{conc}

\begin{table*}
\caption{The assembly of stars in massive galaxies \label{tbl-1}}
\centering
\begin{tabular}{c | c c }
\hline\hline
  & \multicolumn{1}{c}{In-situ} & \multicolumn{1}{c}{Accreted} \\
\hline
{\bf Epoch}  & $ 6 \gtrsim z \gtrsim 2$ & $3 \gtrsim z > 0$ \\
{\bf Baryonic mass source}  & cold gas flows & minor \& major mergers \\
{\bf Size of region}  &  $r_{1/2} \approx  2 \rm kpc $  & $ r_{1/2} \approx  7 \rm kpc $ \\
{\bf Energetics} & Dissipational & Conservative  \\
\hline
\end{tabular}
\label{tab:phases}
\end{table*}

We present results from 39 cosmological re-simulations of 
dark matter halos including gas and star formation covering a mass range of almost two orders of 
magnitude in virial mass. In the study presented here we used the simulations to investigate 
fundamental formation and assembly processes, i.e. how and when do galaxies get their gas and stars, 
and how does this influence the present day galaxy properties. 

We have shown that it can be useful, at a very basic level, to distinguish between stars 
that are created inside the galaxies themselves (in-situ) and those that are created outside 
(ex-situ) and are accreted later on. The division into these two separate phases is quite clean 
(see Fig. \ref{in-situ_mult}) with the in-situ stars typically formed closer 
than $r/r_{vir} \sim 10^{-1}$, i.e. within the galaxy, and the ex-situ stars formed outside the galaxy 
at $r/r_{vir} \sim 10^{0.5}$ to $10$. Independent of galaxy mass we find that the formation of the 
accreted stars peaks at redshift $z \approx 4$. The in-situ star formation as well has an early peak 
but extends over a longer period of time. The ratio of stars that are created in-situ to the accreted 
stars, however varies strongly for galaxies of different masses. We find that for massive galaxies 
$(\sim 1.9 - 3.6 \times 10^{11} h^{-1}M_{\odot})$ the 
contribution of in-situ and accreted stars becomes comparable early $( z \approx 2)$ 
and the accreted stars can account for up to 87\% of the final stellar mass. The lower mass galaxies 
$(\sim 4 - 10 \times 10^{10} h^{-1}M_{\odot})$ 
still can have a high fraction of in-situ formed stars up to 60\% at the present day. They show a significant 
amount of in-situ star formation throughout the whole simulation time. The large difference in time when 
those accreted stars are actually formed and when they are finally 
assimilated by their host, together with the trend shown of the ratio of in-situ formed stars explains 
the phenomenon of 'downsizing' (see also \citep{2006MNRAS.366..499D} for semi-analytical simulations). 
The more massive galaxies consist mainly of accreted and therefore old stars leading to the dependence 
of mean stellar age to galaxy mass shown in Fig. \ref{fig:gmass-age}. The massive galaxies in our 
sample assemble about half their mass below a redshift of z=1. This mass increase, caused by stellar accretion
and merging, is not accompanied by significant star formation, can be a significant contribution to 
the observed increase of stellar mass in the early-type galaxy population since z=1 (see e.g. 
\citealp{2007ApJ...654..858B,2007ApJ...665..265F}).

We find that the accreted stars are primarily responsible for the low redshift size increase in massive 
galaxies (see e.g. \citealp{2009MNRAS.394.1978H}). 
When looking at the half-mass radii of the galaxies and the half-mass radii of the in-situ created 
and accreted components, we find that the half-mass radius of the in-situ created stars is only 
weakly dependent on the galaxy mass and is quite small $(\lesssim 3 \rm kpc \, h^{-1})$. This component forms at redshift
$ z >2$ and makes the compact cores of the galaxies (see e.g. \citealp{2008ApJ...677L...5V} and references therein).
The larger sizes of galaxies with larger mass are mainly due to the accreted stars creating an outer envelope with 
half-mass radii exceeding $8 \rm kpc \, h^{-1}$.

Our simulations overestimate the stellar mass of the galaxies by roughly a factor of 2. This is 
probably due to the lack of ejective and preventive feedback mechanism in our simulations. 
The stars that are accreted as well as the early formed in-situ stars are generated in small systems where winds are most 
effective and lead to lower star formation rates and therefore lower accretion rates at lower redshifts. The late in-situ star
formation should be diminished by AGN feedback particularly in the massive systems. 
It will be worthwhile to investigate
whether and how the inclusion of those processes could influence the presented balance of in-situ star formation to
stellar accretion.

The description of galaxy formation as a two phase process followed in a seemingly natural way from our 
detailed hydro simulations and is organized into a coherent scheme in Table \ref{tab:phases}. It is not 
intended as a rival to other ways of seeing galaxy formation but 
rather as a framework within which the physical processes can be understood in a straightforward way. 
Early, in-situ star formation is clearly similar to that resulting from the 'cold flow' picture \citep{2009Natur.457..451D,2005MNRAS.363....2K}
or the earlier descriptive term 'dissipative collapse'. In fact the in-situ 
phase bears an uncanny resemblance to the 'monolithic collapse' model \citep{1962ApJ...136..748E,1967ApJ...147..868P,1969MNRAS.145..405L,1973ApJ...179..427S,1975MNRAS.173..671L}. The late assembly phase of massive galaxies has many aspects 
similar to the 'dry merger' paradigm investigated by many authors \citep{2003ApJ...597L.117K,2006MNRAS.370..902K,
2006ApJ...636L..81N,2009ApJ...698.1232V,2009ApJ...697.1290B,2009ApJ...703.1531N} with the added qualification 
that most of the accreted stellar systems are low in mass compared to the final assembled i.e. minor mergers 
dominate. There appears to be recent archaeological \citep{2010MNRAS.tmpL..97C} and direct observational for 
this scenario. \citet{2010ApJ...709.1018V} conclude that massive compact galaxies at z=2 (the end of the in-situ phase) 
have increased their mass at radii $r >5kpc$ by a factor of $\approx$ 4 since z=2 with the mass at smaller 
radii being essentially unchanged.

\begin{acknowledgements}
We thank the referee for valuable comments on the paper. We are grateful to S. D. M. White, S. Faber, P. van Dokkum, M. Kriek
, C. Conroy and M. Franx for helpful discussions on the manuscript.
The simulation were partly performed at the Princeton PICSciE HPC center. 
This research was supported by the DFG cluster of excellence 'Origin and Structure of the Universe'. 
\end{acknowledgements}

%\nocite{*}
%\bibliography{/Users/naab/WORK/PAPER/REFERENCES/references.bib}
\bibliography{references}

\begin{thebibliography}{114}
\expandafter\ifx\csname natexlab\endcsname\relax\def\natexlab#1{#1}\fi

\bibitem[{{Abadi} {et~al.}(2010){Abadi}, {Navarro}, {Fardal}, {Babul}, \&
  {Steinmetz}}]{2010MNRAS.tmp..847A}
{Abadi}, M.~G., {Navarro}, J.~F., {Fardal}, M., {Babul}, A., \& {Steinmetz}, M.
  2010, \mnras, 847

\bibitem[{{Abadi} {et~al.}(2003){Abadi}, {Navarro}, {Steinmetz}, \&
  {Eke}}]{2003ApJ...597...21A}
{Abadi}, M.~G., {Navarro}, J.~F., {Steinmetz}, M., \& {Eke}, V.~R. 2003, \apj,
  597, 21

\bibitem[{{Agertz} {et~al.}(2010){Agertz}, {Teyssier}, \&
  {Moore}}]{2010arXiv1004.0005A}
{Agertz}, O., {Teyssier}, R., \& {Moore}, B. 2010, ArXiv e-prints, 1004.0005

\bibitem[{{Auger} {et~al.}(2010){Auger}, {Treu}, {Gavazzi}, {Bolton},
  {Koopmans}, \& {Marshall}}]{2010arXiv1007.2409A}
{Auger}, M.~W., {Treu}, T., {Gavazzi}, R., {Bolton}, A.~S., {Koopmans},
  L.~V.~E., \& {Marshall}, P.~J. 2010, ArXiv e-prints, 1007.2409

\bibitem[{{Behroozi} {et~al.}(2010){Behroozi}, {Conroy}, \&
  {Wechsler}}]{2010ApJ...717..379B}
{Behroozi}, P.~S., {Conroy}, C., \& {Wechsler}, R.~H. 2010, \apj, 717, 379

\bibitem[{{Bender} {et~al.}(1992){Bender}, {Burstein}, \&
  {Faber}}]{1992ApJ...399..462B}
{Bender}, R., {Burstein}, D., \& {Faber}, S.~M. 1992, \apj, 399, 462

\bibitem[{{Bertschinger}(1995)}]{1995astro.ph..6070B}
{Bertschinger}, E. 1995, ArXiv Astrophysics e-prints, 9506070

\bibitem[{{Bertschinger}(2001)}]{2001ApJS..137....1B}
---. 2001, \apjs, 137, 1

\bibitem[{{Bezanson} {et~al.}(2009){Bezanson}, {van Dokkum}, {Tal},
  {Marchesini}, {Kriek}, {Franx}, \& {Coppi}}]{2009ApJ...697.1290B}
{Bezanson}, R., {van Dokkum}, P.~G., {Tal}, T., {Marchesini}, D., {Kriek}, M.,
  {Franx}, M., \& {Coppi}, P. 2009, \apj, 697, 1290

\bibitem[{{Blumenthal} {et~al.}(1986){Blumenthal}, {Faber}, {Flores}, \&
  {Primack}}]{1986ApJ...301...27B}
{Blumenthal}, G.~R., {Faber}, S.~M., {Flores}, R., \& {Primack}, J.~R. 1986,
  \apj, 301, 27

\bibitem[{{Brown} {et~al.}(2007){Brown}, {Dey}, {Jannuzi}, {Brand}, {Benson},
  {Brodwin}, {Croton}, \& {Eisenhardt}}]{2007ApJ...654..858B}
{Brown}, M.~J.~I., {Dey}, A., {Jannuzi}, B.~T., {Brand}, K., {Benson}, A.~J.,
  {Brodwin}, M., {Croton}, D.~J., \& {Eisenhardt}, P.~R. 2007, \apj, 654, 858

\bibitem[{{Bullock} {et~al.}(2001){Bullock}, {Dekel}, {Kolatt}, {Kravtsov},
  {Klypin}, {Porciani}, \& {Primack}}]{2001ApJ...555..240B}
{Bullock}, J.~S., {Dekel}, A., {Kolatt}, T.~S., {Kravtsov}, A.~V., {Klypin},
  A.~A., {Porciani}, C., \& {Primack}, J.~R. 2001, \apj, 555, 240

\bibitem[{{Cen} \& {Chisari}(2010)}]{2010arXiv1005.1451C}
{Cen}, R. \& {Chisari}, N.~E. 2010, ArXiv e-prints, 1005.1451

\bibitem[{{Cen} \& {Ostriker}(1999)}]{1999ApJ...514....1C}
{Cen}, R. \& {Ostriker}, J.~P. 1999, \apj, 514, 1

\bibitem[{{Coccato} {et~al.}(2010){Coccato}, {Gerhard}, \&
  {Arnaboldi}}]{2010MNRAS.tmpL..97C}
{Coccato}, L., {Gerhard}, O., \& {Arnaboldi}, M. 2010, \mnras, L97+

\bibitem[{{Conroy} \& {Wechsler}(2009)}]{2009ApJ...696..620C}
{Conroy}, C. \& {Wechsler}, R.~H. 2009, \apj, 696, 620

\bibitem[{{Croton} {et~al.}(2006){Croton}, {Springel}, {White}, {De Lucia},
  {Frenk}, {Gao}, {Jenkins}, {Kauffmann}, {Navarro}, \&
  {Yoshida}}]{2006MNRAS.365...11C}
{Croton}, D.~J., {Springel}, V., {White}, S.~D.~M., {De Lucia}, G., {Frenk},
  C.~S., {Gao}, L., {Jenkins}, A., {Kauffmann}, G., {Navarro}, J.~F., \&
  {Yoshida}, N. 2006, \mnras, 365, 11

\bibitem[{{Dav{\'e}} {et~al.}(1999){Dav{\'e}}, {Hernquist}, {Katz}, \&
  {Weinberg}}]{1999ApJ...511..521D}
{Dav{\'e}}, R., {Hernquist}, L., {Katz}, N., \& {Weinberg}, D.~H. 1999, \apj,
  511, 521

\bibitem[{{De Lucia} {et~al.}(2006){De Lucia}, {Springel}, {White}, {Croton},
  \& {Kauffmann}}]{2006MNRAS.366..499D}
{De Lucia}, G., {Springel}, V., {White}, S.~D.~M., {Croton}, D., \&
  {Kauffmann}, G. 2006, \mnras, 366, 499

\bibitem[{{Debattista} {et~al.}(2008){Debattista}, {Moore}, {Quinn},
  {Kazantzidis}, {Maas}, {Mayer}, {Read}, \& {Stadel}}]{2008ApJ...681.1076D}
{Debattista}, V.~P., {Moore}, B., {Quinn}, T., {Kazantzidis}, S., {Maas}, R.,
  {Mayer}, L., {Read}, J., \& {Stadel}, J. 2008, \apj, 681, 1076

\bibitem[{{Dehnen}(2001)}]{2001MNRAS.324..273D}
{Dehnen}, W. 2001, \mnras, 324, 273

\bibitem[{{Dekel} \& {Birnboim}(2006)}]{2006MNRAS.368....2D}
{Dekel}, A. \& {Birnboim}, Y. 2006, \mnras, 368, 2

\bibitem[{{Dekel} {et~al.}(2009{\natexlab{a}}){Dekel}, {Birnboim}, {Engel},
  {Freundlich}, {Goerdt}, {Mumcuoglu}, {Neistein}, {Pichon}, {Teyssier}, \&
  {Zinger}}]{2009Natur.457..451D}
{Dekel}, A., {Birnboim}, Y., {Engel}, G., {Freundlich}, J., {Goerdt}, T.,
  {Mumcuoglu}, M., {Neistein}, E., {Pichon}, C., {Teyssier}, R., \& {Zinger},
  E. 2009{\natexlab{a}}, \nat, 457, 451

\bibitem[{{Dekel} {et~al.}(2009{\natexlab{b}}){Dekel}, {Sari}, \&
  {Ceverino}}]{2009ApJ...703..785D}
{Dekel}, A., {Sari}, R., \& {Ceverino}, D. 2009{\natexlab{b}}, \apj, 703, 785

\bibitem[{{Dekel} \& {Silk}(1986)}]{1986ApJ...303...39D}
{Dekel}, A. \& {Silk}, J. 1986, \apj, 303, 39

\bibitem[{{Di Matteo} {et~al.}(2008){Di Matteo}, {Colberg}, {Springel},
  {Hernquist}, \& {Sijacki}}]{2008ApJ...676...33D}
{Di Matteo}, T., {Colberg}, J., {Springel}, V., {Hernquist}, L., \& {Sijacki},
  D. 2008, \apj, 676, 33

\bibitem[{{Dubinski}(1994)}]{1994ApJ...431..617D}
{Dubinski}, J. 1994, \apj, 431, 617

\bibitem[{{Eggen} {et~al.}(1962){Eggen}, {Lynden-Bell}, \&
  {Sandage}}]{1962ApJ...136..748E}
{Eggen}, O.~J., {Lynden-Bell}, D., \& {Sandage}, A.~R. 1962, \apj, 136, 748

\bibitem[{{Faber} {et~al.}(2007){Faber}, {Willmer}, {Wolf}, {Koo}, {Weiner},
  {Newman}, {Im}, {Coil}, {Conroy}, {Cooper}, {Davis}, {Finkbeiner}, {Gerke},
  {Gebhardt}, {Groth}, {Guhathakurta}, {Harker}, {Kaiser}, {Kassin},
  {Kleinheinrich}, {Konidaris}, {Kron}, {Lin}, {Luppino}, {Madgwick},
  {Meisenheimer}, {Noeske}, {Phillips}, {Sarajedini}, {Schiavon}, {Simard},
  {Szalay}, {Vogt}, \& {Yan}}]{2007ApJ...665..265F}
{Faber}, S.~M., {Willmer}, C.~N.~A., {Wolf}, C., {Koo}, D.~C., {Weiner}, B.~J.,
  {Newman}, J.~A., {Im}, M., {Coil}, A.~L., {Conroy}, C., {Cooper}, M.~C.,
  {Davis}, M., {Finkbeiner}, D.~P., {Gerke}, B.~F., {Gebhardt}, K., {Groth},
  E.~J., {Guhathakurta}, P., {Harker}, J., {Kaiser}, N., {Kassin}, S.,
  {Kleinheinrich}, M., {Konidaris}, N.~P., {Kron}, R.~G., {Lin}, L., {Luppino},
  G., {Madgwick}, D.~S., {Meisenheimer}, K., {Noeske}, K.~G., {Phillips},
  A.~C., {Sarajedini}, V.~L., {Schiavon}, R.~P., {Simard}, L., {Szalay}, A.~S.,
  {Vogt}, N.~P., \& {Yan}, R. 2007, \apj, 665, 265

\bibitem[{{Feldmann} {et~al.}(2010){Feldmann}, {Carollo}, {Mayer}, {Renzini},
  {Lake}, {Quinn}, {Stinson}, \& {Yepes}}]{2010ApJ...709..218F}
{Feldmann}, R., {Carollo}, C.~M., {Mayer}, L., {Renzini}, A., {Lake}, G.,
  {Quinn}, T., {Stinson}, G.~S., \& {Yepes}, G. 2010, \apj, 709, 218

\bibitem[{{F{\"o}rster Schreiber} {et~al.}(2009){F{\"o}rster Schreiber},
  {Genzel}, {Bouch{\'e}}, {Cresci}, {Davies}, {Buschkamp}, {Shapiro},
  {Tacconi}, {Hicks}, {Genel}, {Shapley}, {Erb}, {Steidel}, {Lutz},
  {Eisenhauer}, {Gillessen}, {Sternberg}, {Renzini}, {Cimatti}, {Daddi},
  {Kurk}, {Lilly}, {Kong}, {Lehnert}, {Nesvadba}, {Verma}, {McCracken},
  {Arimoto}, {Mignoli}, \& {Onodera}}]{2009ApJ...706.1364F}
{F{\"o}rster Schreiber}, N.~M., {Genzel}, R., {Bouch{\'e}}, N., {Cresci}, G.,
  {Davies}, R., {Buschkamp}, P., {Shapiro}, K., {Tacconi}, L.~J., {Hicks},
  E.~K.~S., {Genel}, S., {Shapley}, A.~E., {Erb}, D.~K., {Steidel}, C.~C.,
  {Lutz}, D., {Eisenhauer}, F., {Gillessen}, S., {Sternberg}, A., {Renzini},
  A., {Cimatti}, A., {Daddi}, E., {Kurk}, J., {Lilly}, S., {Kong}, X.,
  {Lehnert}, M.~D., {Nesvadba}, N., {Verma}, A., {McCracken}, H., {Arimoto},
  N., {Mignoli}, M., \& {Onodera}, M. 2009, \apj, 706, 1364

\bibitem[{{F{\"o}rster Schreiber} {et~al.}(2006){F{\"o}rster Schreiber},
  {Genzel}, {Lehnert}, {Bouch{\'e}}, {Verma}, {Erb}, {Shapley}, {Steidel},
  {Davies}, {Lutz}, {Nesvadba}, {Tacconi}, {Eisenhauer}, {Abuter}, {Gilbert},
  {Gillessen}, \& {Sternberg}}]{2006ApJ...645.1062F}
{F{\"o}rster Schreiber}, N.~M., {Genzel}, R., {Lehnert}, M.~D., {Bouch{\'e}},
  N., {Verma}, A., {Erb}, D.~K., {Shapley}, A.~E., {Steidel}, C.~C., {Davies},
  R., {Lutz}, D., {Nesvadba}, N., {Tacconi}, L.~J., {Eisenhauer}, F., {Abuter},
  R., {Gilbert}, A., {Gillessen}, S., \& {Sternberg}, A. 2006, \apj, 645, 1062

\bibitem[{{Franx} {et~al.}(2008){Franx}, {van Dokkum}, {Schreiber}, {Wuyts},
  {Labb{\'e}}, \& {Toft}}]{2008ApJ...688..770F}
{Franx}, M., {van Dokkum}, P.~G., {Schreiber}, N.~M.~F., {Wuyts}, S.,
  {Labb{\'e}}, I., \& {Toft}, S. 2008, \apj, 688, 770

\bibitem[{{Genzel} {et~al.}(2006){Genzel}, {Tacconi}, {Eisenhauer},
  {F{\"o}rster Schreiber}, {Cimatti}, {Daddi}, {Bouch{\'e}}, {Davies},
  {Lehnert}, {Lutz}, {Nesvadba}, {Verma}, {Abuter}, {Shapiro}, {Sternberg},
  {Renzini}, {Kong}, {Arimoto}, \& {Mignoli}}]{2006Natur.442..786G}
{Genzel}, R., {Tacconi}, L.~J., {Eisenhauer}, F., {F{\"o}rster Schreiber},
  N.~M., {Cimatti}, A., {Daddi}, E., {Bouch{\'e}}, N., {Davies}, R., {Lehnert},
  M.~D., {Lutz}, D., {Nesvadba}, N., {Verma}, A., {Abuter}, R., {Shapiro}, K.,
  {Sternberg}, A., {Renzini}, A., {Kong}, X., {Arimoto}, N., \& {Mignoli}, M.
  2006, \nat, 442, 786

\bibitem[{{Gnedin} {et~al.}(2004){Gnedin}, {Kravtsov}, {Klypin}, \&
  {Nagai}}]{2004ApJ...616...16G}
{Gnedin}, O.~Y., {Kravtsov}, A.~V., {Klypin}, A.~A., \& {Nagai}, D. 2004, \apj,
  616, 16

\bibitem[{{Governato} {et~al.}(2010){Governato}, {Brook}, {Mayer}, {Brooks},
  {Rhee}, {Wadsley}, {Jonsson}, {Willman}, {Stinson}, {Quinn}, \&
  {Madau}}]{2010Natur.463..203G}
{Governato}, F., {Brook}, C., {Mayer}, L., {Brooks}, A., {Rhee}, G., {Wadsley},
  J., {Jonsson}, P., {Willman}, B., {Stinson}, G., {Quinn}, T., \& {Madau}, P.
  2010, \nat, 463, 203

\bibitem[{{Governato} {et~al.}(2009){Governato}, {Brook}, {Brooks}, {Mayer},
  {Willman}, {Jonsson}, {Stilp}, {Pope}, {Christensen}, {Wadsley}, \&
  {Quinn}}]{2009MNRAS.398..312G}
{Governato}, F., {Brook}, C.~B., {Brooks}, A.~M., {Mayer}, L., {Willman}, B.,
  {Jonsson}, P., {Stilp}, A.~M., {Pope}, L., {Christensen}, C., {Wadsley}, J.,
  \& {Quinn}, T. 2009, \mnras, 398, 312

\bibitem[{{Governato} {et~al.}(2007){Governato}, {Willman}, {Mayer}, {Brooks},
  {Stinson}, {Valenzuela}, {Wadsley}, \& {Quinn}}]{2007MNRAS.374.1479G}
{Governato}, F., {Willman}, B., {Mayer}, L., {Brooks}, A., {Stinson}, G.,
  {Valenzuela}, O., {Wadsley}, J., \& {Quinn}, T. 2007, \mnras, 374, 1479

\bibitem[{{Grossi} \& {Springel}(2009)}]{2009MNRAS.394.1559G}
{Grossi}, M. \& {Springel}, V. 2009, \mnras, 394, 1559

\bibitem[{{Guo} {et~al.}(2010{\natexlab{a}}){Guo}, {White}, {Boylan-Kolchin},
  {De Lucia}, {Kauffmann}, {Lemson}, {Li}, {Springel}, \&
  {Weinmann}}]{2010arXiv1006.0106G}
{Guo}, Q., {White}, S., {Boylan-Kolchin}, M., {De Lucia}, G., {Kauffmann}, G.,
  {Lemson}, G., {Li}, C., {Springel}, V., \& {Weinmann}, S. 2010{\natexlab{a}},
  ArXiv e-prints

\bibitem[{{Guo} {et~al.}(2010{\natexlab{b}}){Guo}, {White}, {Li}, \&
  {Boylan-Kolchin}}]{2010MNRAS.404.1111G}
{Guo}, Q., {White}, S., {Li}, C., \& {Boylan-Kolchin}, M. 2010{\natexlab{b}},
  \mnras, 404, 1111

\bibitem[{{Haardt} \& {Madau}(1996)}]{1996ApJ...461...20H}
{Haardt}, F. \& {Madau}, P. 1996, \apj, 461, 20

\bibitem[{{Hambrick} {et~al.}(2010){Hambrick}, {Ostriker}, {Johansson}, \&
  {Naab}}]{2010arXiv1009.6005H}
{Hambrick}, D.~C., {Ostriker}, J.~P., {Johansson}, P.~H., \& {Naab}, T. 2010,
  ArXiv e-prints, 1009.6005

\bibitem[{{Hopkins} {et~al.}(2009){Hopkins}, {Bundy}, {Croton}, {Hernquist},
  {Keres}, {Khochfar}, {Stewart}, {Wetzel}, \& {Younger}}]{2009arXiv0906.5357H}
{Hopkins}, P.~F., {Bundy}, K., {Croton}, D., {Hernquist}, L., {Keres}, D.,
  {Khochfar}, S., {Stewart}, K., {Wetzel}, A., \& {Younger}, J.~D. 2009, ArXiv
  e-prints, 0906.5357

\bibitem[{{Hopkins} {et~al.}(2010){Hopkins}, {Bundy}, {Hernquist}, {Wuyts}, \&
  {Cox}}]{2010MNRAS.401.1099H}
{Hopkins}, P.~F., {Bundy}, K., {Hernquist}, L., {Wuyts}, S., \& {Cox}, T.~J.
  2010, \mnras, 401, 1099

\bibitem[{{Hyde} \& {Bernardi}(2009)}]{2009MNRAS.394.1978H}
{Hyde}, J.~B. \& {Bernardi}, M. 2009, \mnras, 394, 1978

\bibitem[{{Jenkins}(2010)}]{2010MNRAS.403.1859J}
{Jenkins}, A. 2010, \mnras, 403, 1859

\bibitem[{{Jesseit} {et~al.}(2002){Jesseit}, {Naab}, \&
  {Burkert}}]{2002ApJ...571L..89J}
{Jesseit}, R., {Naab}, T., \& {Burkert}, A. 2002, \apjl, 571, L89

\bibitem[{{Johansson} \& {Efstathiou}(2006)}]{2006MNRAS.371.1519J}
{Johansson}, P.~H. \& {Efstathiou}, G. 2006, \mnras, 371, 1519

\bibitem[{{Johansson} {et~al.}(2009){Johansson}, {Naab}, \&
  {Ostriker}}]{2009ApJ...697L..38J}
{Johansson}, P.~H., {Naab}, T., \& {Ostriker}, J.~P. 2009, \apjl, 697, L38

\bibitem[{{Joung} {et~al.}(2009){Joung}, {Cen}, \&
  {Bryan}}]{2009ApJ...692L...1J}
{Joung}, M.~R., {Cen}, R., \& {Bryan}, G.~L. 2009, \apjl, 692, L1

\bibitem[{{Juneau} {et~al.}(2005){Juneau}, {Glazebrook}, {Crampton},
  {McCarthy}, {Savaglio}, {Abraham}, {Carlberg}, {Chen}, {Le Borgne}, {Marzke},
  {Roth}, {J{\o}rgensen}, {Hook}, \& {Murowinski}}]{2005ApJ...619L.135J}
{Juneau}, S., {Glazebrook}, K., {Crampton}, D., {McCarthy}, P.~J., {Savaglio},
  S., {Abraham}, R., {Carlberg}, R.~G., {Chen}, H., {Le Borgne}, D., {Marzke},
  R.~O., {Roth}, K., {J{\o}rgensen}, I., {Hook}, I., \& {Murowinski}, R. 2005,
  \apjl, 619, L135

\bibitem[{{Katz} {et~al.}(1996){Katz}, {Weinberg}, \&
  {Hernquist}}]{1996ApJS..105...19K}
{Katz}, N., {Weinberg}, D.~H., \& {Hernquist}, L. 1996, \apjs, 105, 19

\bibitem[{{Kauffmann} {et~al.}(1999){Kauffmann}, {Colberg}, {Diaferio}, \&
  {White}}]{1999MNRAS.303..188K}
{Kauffmann}, G., {Colberg}, J.~M., {Diaferio}, A., \& {White}, S.~D.~M. 1999,
  \mnras, 303, 188

\bibitem[{{Kere{\v s}} {et~al.}(2009{\natexlab{a}}){Kere{\v s}}, {Katz},
  {Dav{\'e}}, {Fardal}, \& {Weinberg}}]{2009MNRAS.396.2332K}
{Kere{\v s}}, D., {Katz}, N., {Dav{\'e}}, R., {Fardal}, M., \& {Weinberg},
  D.~H. 2009{\natexlab{a}}, \mnras, 396, 2332

\bibitem[{{Kere{\v s}} {et~al.}(2009{\natexlab{b}}){Kere{\v s}}, {Katz},
  {Fardal}, {Dav{\'e}}, \& {Weinberg}}]{2009MNRAS.395..160K}
{Kere{\v s}}, D., {Katz}, N., {Fardal}, M., {Dav{\'e}}, R., \& {Weinberg},
  D.~H. 2009{\natexlab{b}}, \mnras, 395, 160

\bibitem[{{Kere{\v s}} {et~al.}(2005){Kere{\v s}}, {Katz}, {Weinberg}, \&
  {Dav{\'e}}}]{2005MNRAS.363....2K}
{Kere{\v s}}, D., {Katz}, N., {Weinberg}, D.~H., \& {Dav{\'e}}, R. 2005,
  \mnras, 363, 2

\bibitem[{{Khochfar} \& {Burkert}(2003)}]{2003ApJ...597L.117K}
{Khochfar}, S. \& {Burkert}, A. 2003, \apjl, 597, L117

\bibitem[{{Khochfar} \& {Silk}(2006{\natexlab{a}})}]{2006ApJ...648L..21K}
{Khochfar}, S. \& {Silk}, J. 2006{\natexlab{a}}, \apjl, 648, L21

\bibitem[{{Khochfar} \& {Silk}(2006{\natexlab{b}})}]{2006MNRAS.370..902K}
---. 2006{\natexlab{b}}, \mnras, 370, 902

\bibitem[{{Komatsu} {et~al.}(2010){Komatsu}, {Smith}, {Dunkley}, {Bennett},
  {Gold}, {Hinshaw}, {Jarosik}, {Larson}, {Nolta}, {Page}, {Spergel},
  {Halpern}, {Hill}, {Kogut}, {Limon}, {Meyer}, {Odegard}, {Tucker}, {Weiland},
  {Wollack}, \& {Wright}}]{2010arXiv1001.4538K}
{Komatsu}, E., {Smith}, K.~M., {Dunkley}, J., {Bennett}, C.~L., {Gold}, B.,
  {Hinshaw}, G., {Jarosik}, N., {Larson}, D., {Nolta}, M.~R., {Page}, L.,
  {Spergel}, D.~N., {Halpern}, M., {Hill}, R.~S., {Kogut}, A., {Limon}, M.,
  {Meyer}, S.~S., {Odegard}, N., {Tucker}, G.~S., {Weiland}, J.~L., {Wollack},
  E., \& {Wright}, E.~L. 2010, ArXiv e-prints, 1001.4538

\bibitem[{{Kriek} {et~al.}(2008){Kriek}, {van der Wel}, {van Dokkum}, {Franx},
  \& {Illingworth}}]{2008ApJ...682..896K}
{Kriek}, M., {van der Wel}, A., {van Dokkum}, P.~G., {Franx}, M., \&
  {Illingworth}, G.~D. 2008, \apj, 682, 896

\bibitem[{{Lackner} \& {Ostriker}(2010)}]{2010ApJ...712...88L}
{Lackner}, C.~N. \& {Ostriker}, J.~P. 2010, \apj, 712, 88

\bibitem[{{Larson}(1969)}]{1969MNRAS.145..405L}
{Larson}, R.~B. 1969, \mnras, 145, 405

\bibitem[{{Larson}(1974)}]{1974MNRAS.169..229L}
---. 1974, \mnras, 169, 229

\bibitem[{{Larson}(1975)}]{1975MNRAS.173..671L}
---. 1975, \mnras, 173, 671

\bibitem[{{Mandelbaum} {et~al.}(2006){Mandelbaum}, {Seljak}, {Kauffmann},
  {Hirata}, \& {Brinkmann}}]{2006MNRAS.368..715M}
{Mandelbaum}, R., {Seljak}, U., {Kauffmann}, G., {Hirata}, C.~M., \&
  {Brinkmann}, J. 2006, \mnras, 368, 715

\bibitem[{{Marchesini} {et~al.}(2009){Marchesini}, {van Dokkum}, {F{\"o}rster
  Schreiber}, {Franx}, {Labb{\'e}}, \& {Wuyts}}]{2009ApJ...701.1765M}
{Marchesini}, D., {van Dokkum}, P.~G., {F{\"o}rster Schreiber}, N.~M., {Franx},
  M., {Labb{\'e}}, I., \& {Wuyts}, S. 2009, \apj, 701, 1765

\bibitem[{{McCarthy} {et~al.}(2010){McCarthy}, {Schaye}, {Ponman}, {Bower},
  {Booth}, {Dalla Vecchia}, {Crain}, {Springel}, {Theuns}, \&
  {Wiersma}}]{2010MNRAS.tmp..740M}
{McCarthy}, I.~G., {Schaye}, J., {Ponman}, T.~J., {Bower}, R.~G., {Booth},
  C.~M., {Dalla Vecchia}, C., {Crain}, R.~A., {Springel}, V., {Theuns}, T., \&
  {Wiersma}, R.~P.~C. 2010, \mnras, 740

\bibitem[{{McKee} \& {Ostriker}(1977)}]{1977ApJ...218..148M}
{McKee}, C.~F. \& {Ostriker}, J.~P. 1977, \apj, 218, 148

\bibitem[{{Meza} {et~al.}(2005){Meza}, {Navarro}, {Abadi}, \&
  {Steinmetz}}]{2005MNRAS.359...93M}
{Meza}, A., {Navarro}, J.~F., {Abadi}, M.~G., \& {Steinmetz}, M. 2005, \mnras,
  359, 93

\bibitem[{{Meza} {et~al.}(2003){Meza}, {Navarro}, {Steinmetz}, \&
  {Eke}}]{2003ApJ...590..619M}
{Meza}, A., {Navarro}, J.~F., {Steinmetz}, M., \& {Eke}, V.~R. 2003, \apj, 590,
  619

\bibitem[{{Monaghan}(1992)}]{1992ARA&A..30..543M}
{Monaghan}, J.~J. 1992, \araa, 30, 543

\bibitem[{{Moster} {et~al.}(2010){Moster}, {Somerville}, {Maulbetsch}, {van den
  Bosch}, {Macci{\`o}}, {Naab}, \& {Oser}}]{2010ApJ...710..903M}
{Moster}, B.~P., {Somerville}, R.~S., {Maulbetsch}, C., {van den Bosch}, F.~C.,
  {Macci{\`o}}, A.~V., {Naab}, T., \& {Oser}, L. 2010, \apj, 710, 903

\bibitem[{{Naab} {et~al.}(2009){Naab}, {Johansson}, \&
  {Ostriker}}]{2009ApJ...699L.178N}
{Naab}, T., {Johansson}, P.~H., \& {Ostriker}, J.~P. 2009, \apjl, 699, L178

\bibitem[{{Naab} {et~al.}(2007){Naab}, {Johansson}, {Ostriker}, \&
  {Efstathiou}}]{2007ApJ...658..710N}
{Naab}, T., {Johansson}, P.~H., {Ostriker}, J.~P., \& {Efstathiou}, G. 2007,
  \apj, 658, 710

\bibitem[{{Naab} {et~al.}(2006){Naab}, {Khochfar}, \&
  {Burkert}}]{2006ApJ...636L..81N}
{Naab}, T., {Khochfar}, S., \& {Burkert}, A. 2006, \apjl, 636, L81

\bibitem[{{Nagamine} {et~al.}(2005){Nagamine}, {Cen}, {Hernquist}, {Ostriker},
  \& {Springel}}]{2005ApJ...627..608N}
{Nagamine}, K., {Cen}, R., {Hernquist}, L., {Ostriker}, J.~P., \& {Springel},
  V. 2005, \apj, 627, 608

\bibitem[{{Nagamine} {et~al.}(2006){Nagamine}, {Ostriker}, {Fukugita}, \&
  {Cen}}]{2006ApJ...653..881N}
{Nagamine}, K., {Ostriker}, J.~P., {Fukugita}, M., \& {Cen}, R. 2006, \apj,
  653, 881

\bibitem[{{Navarro} {et~al.}(1997){Navarro}, {Frenk}, \&
  {White}}]{1997ApJ...490..493N}
{Navarro}, J.~F., {Frenk}, C.~S., \& {White}, S.~D.~M. 1997, \apj, 490, 493

\bibitem[{{Nelan} {et~al.}(2005){Nelan}, {Smith}, {Hudson}, {Wegner}, {Lucey},
  {Moore}, {Quinney}, \& {Suntzeff}}]{2005ApJ...632..137N}
{Nelan}, J.~E., {Smith}, R.~J., {Hudson}, M.~J., {Wegner}, G.~A., {Lucey},
  J.~R., {Moore}, S.~A.~W., {Quinney}, S.~J., \& {Suntzeff}, N.~B. 2005, \apj,
  632, 137

\bibitem[{{Nipoti} {et~al.}(2009{\natexlab{a}}){Nipoti}, {Treu}, {Auger}, \&
  {Bolton}}]{2009ApJ...706L..86N}
{Nipoti}, C., {Treu}, T., {Auger}, M.~W., \& {Bolton}, A.~S.
  2009{\natexlab{a}}, \apjl, 706, L86

\bibitem[{{Nipoti} {et~al.}(2009{\natexlab{b}}){Nipoti}, {Treu}, \&
  {Bolton}}]{2009ApJ...703.1531N}
{Nipoti}, C., {Treu}, T., \& {Bolton}, A.~S. 2009{\natexlab{b}}, \apj, 703,
  1531

\bibitem[{{Oppenheimer} \& {Dav{\'e}}(2008)}]{2008MNRAS.387..577O}
{Oppenheimer}, B.~D. \& {Dav{\'e}}, R. 2008, \mnras, 387, 577

\bibitem[{{Oppenheimer} {et~al.}(2010){Oppenheimer}, {Dav{\'e}}, {Kere{\v s}},
  {Fardal}, {Katz}, {Kollmeier}, \& {Weinberg}}]{2010MNRAS.tmp..860O}
{Oppenheimer}, B.~D., {Dav{\'e}}, R., {Kere{\v s}}, D., {Fardal}, M., {Katz},
  N., {Kollmeier}, J.~A., \& {Weinberg}, D.~H. 2010, \mnras, 860

\bibitem[{{Partridge} \& {Peebles}(1967)}]{1967ApJ...147..868P}
{Partridge}, R.~B. \& {Peebles}, P.~J.~E. 1967, \apj, 147, 868

\bibitem[{{Pettini} {et~al.}(2001){Pettini}, {Shapley}, {Steidel}, {Cuby},
  {Dickinson}, {Moorwood}, {Adelberger}, \& {Giavalisco}}]{2001ApJ...554..981P}
{Pettini}, M., {Shapley}, A.~E., {Steidel}, C.~C., {Cuby}, J., {Dickinson}, M.,
  {Moorwood}, A.~F.~M., {Adelberger}, K.~L., \& {Giavalisco}, M. 2001, \apj,
  554, 981

\bibitem[{{Piontek} \& {Steinmetz}(2009)}]{2009arXiv0909.4167P}
{Piontek}, F. \& {Steinmetz}, M. 2009, ArXiv e-prints, 0909.4167

\bibitem[{{Power} {et~al.}(2003){Power}, {Navarro}, {Jenkins}, {Frenk},
  {White}, {Springel}, {Stadel}, \& {Quinn}}]{2003MNRAS.338...14P}
{Power}, C., {Navarro}, J.~F., {Jenkins}, A., {Frenk}, C.~S., {White},
  S.~D.~M., {Springel}, V., {Stadel}, J., \& {Quinn}, T. 2003, \mnras, 338, 14

\bibitem[{{Sawala} {et~al.}(2010){Sawala}, {Scannapieco}, {Maio}, \&
  {White}}]{2010MNRAS.402.1599S}
{Sawala}, T., {Scannapieco}, C., {Maio}, U., \& {White}, S. 2010, \mnras, 402,
  1599

\bibitem[{{Scannapieco} {et~al.}(2008){Scannapieco}, {Tissera}, {White}, \&
  {Springel}}]{2008MNRAS.389.1137S}
{Scannapieco}, C., {Tissera}, P.~B., {White}, S.~D.~M., \& {Springel}, V. 2008,
  \mnras, 389, 1137

\bibitem[{{Scannapieco} {et~al.}(2009){Scannapieco}, {White}, {Springel}, \&
  {Tissera}}]{2009MNRAS.396..696S}
{Scannapieco}, C., {White}, S.~D.~M., {Springel}, V., \& {Tissera}, P.~B. 2009,
  \mnras, 396, 696

\bibitem[{{Schaye} {et~al.}(2009){Schaye}, {Dalla Vecchia}, {Booth}, {Wiersma},
  {Theuns}, {Haas}, {Bertone}, {Duffy}, {McCarthy}, \& {van de
  Voort}}]{2009arXiv0909.5196S}
{Schaye}, J., {Dalla Vecchia}, C., {Booth}, C.~M., {Wiersma}, R.~P.~C.,
  {Theuns}, T., {Haas}, M.~R., {Bertone}, S., {Duffy}, A.~R., {McCarthy},
  I.~G., \& {van de Voort}, F. 2009, ArXiv e-prints

\bibitem[{{Schaye} {et~al.}(2010){Schaye}, {Dalla Vecchia}, {Booth}, {Wiersma},
  {Theuns}, {Haas}, {Bertone}, {Duffy}, {McCarthy}, \& {van de
  Voort}}]{2010MNRAS.402.1536S}
---. 2010, \mnras, 402, 0909.5196

\bibitem[{{Searle} {et~al.}(1973){Searle}, {Sargent}, \&
  {Bagnuolo}}]{1973ApJ...179..427S}
{Searle}, L., {Sargent}, W.~L.~W., \& {Bagnuolo}, W.~G. 1973, \apj, 179, 427

\bibitem[{{Sheth} {et~al.}(2001){Sheth}, {Mo}, \&
  {Tormen}}]{2001MNRAS.323....1S}
{Sheth}, R.~K., {Mo}, H.~J., \& {Tormen}, G. 2001, \mnras, 323, 1

\bibitem[{{Somerville} \& {Primack}(1999)}]{1999MNRAS.310.1087S}
{Somerville}, R.~S. \& {Primack}, J.~R. 1999, \mnras, 310, 1087

\bibitem[{{Sommer-Larsen} {et~al.}(2003){Sommer-Larsen}, {G{\"o}tz}, \&
  {Portinari}}]{2003ApJ...596...47S}
{Sommer-Larsen}, J., {G{\"o}tz}, M., \& {Portinari}, L. 2003, \apj, 596, 47

\bibitem[{{Spergel} {et~al.}(2007){Spergel}, {Bean}, {Dor{\'e}}, {Nolta},
  {Bennett}, {Dunkley}, {Hinshaw}, {Jarosik}, {Komatsu}, {Page}, {Peiris},
  {Verde}, {Halpern}, {Hill}, {Kogut}, {Limon}, {Meyer}, {Odegard}, {Tucker},
  {Weiland}, {Wollack}, \& {Wright}}]{2007ApJS..170..377S}
{Spergel}, D.~N., {Bean}, R., {Dor{\'e}}, O., {Nolta}, M.~R., {Bennett}, C.~L.,
  {Dunkley}, J., {Hinshaw}, G., {Jarosik}, N., {Komatsu}, E., {Page}, L.,
  {Peiris}, H.~V., {Verde}, L., {Halpern}, M., {Hill}, R.~S., {Kogut}, A.,
  {Limon}, M., {Meyer}, S.~S., {Odegard}, N., {Tucker}, G.~S., {Weiland},
  J.~L., {Wollack}, E., \& {Wright}, E.~L. 2007, \apjs, 170, 377

\bibitem[{{Springel}(2005)}]{2005MNRAS.364.1105S}
{Springel}, V. 2005, \mnras, 364, 1105

\bibitem[{{Springel} \& {Hernquist}(2002)}]{2002MNRAS.333..649S}
{Springel}, V. \& {Hernquist}, L. 2002, \mnras, 333, 649

\bibitem[{{Springel} \& {Hernquist}(2003)}]{2003MNRAS.339..289S}
---. 2003, \mnras, 339, 289

\bibitem[{{Springel} {et~al.}(2005){Springel}, {White}, {Jenkins}, {Frenk},
  {Yoshida}, {Gao}, {Navarro}, {Thacker}, {Croton}, {Helly}, {Peacock}, {Cole},
  {Thomas}, {Couchman}, {Evrard}, {Colberg}, \& {Pearce}}]{2005Natur.435..629S}
{Springel}, V., {White}, S.~D.~M., {Jenkins}, A., {Frenk}, C.~S., {Yoshida},
  N., {Gao}, L., {Navarro}, J., {Thacker}, R., {Croton}, D., {Helly}, J.,
  {Peacock}, J.~A., {Cole}, S., {Thomas}, P., {Couchman}, H., {Evrard}, A.,
  {Colberg}, J., \& {Pearce}, F. 2005, \nat, 435, 629

\bibitem[{{Steidel} {et~al.}(1999){Steidel}, {Adelberger}, {Giavalisco},
  {Dickinson}, \& {Pettini}}]{1999ApJ...519....1S}
{Steidel}, C.~C., {Adelberger}, K.~L., {Giavalisco}, M., {Dickinson}, M., \&
  {Pettini}, M. 1999, \apj, 519, 1

\bibitem[{{Steidel} {et~al.}(2010){Steidel}, {Erb}, {Shapley}, {Pettini},
  {Reddy}, {Bogosavljevi{\'c}}, {Rudie}, \& {Rakic}}]{2010ApJ...717..289S}
{Steidel}, C.~C., {Erb}, D.~K., {Shapley}, A.~E., {Pettini}, M., {Reddy}, N.,
  {Bogosavljevi{\'c}}, M., {Rudie}, G.~C., \& {Rakic}, O. 2010, \apj, 717, 289

\bibitem[{{Thomas} {et~al.}(2005){Thomas}, {Maraston}, {Bender}, \& {de
  Oliveira}}]{2005ApJ...621..673T}
{Thomas}, D., {Maraston}, C., {Bender}, R., \& {de Oliveira}, C.~M. 2005, \apj,
  621, 673

\bibitem[{{Tiret} {et~al.}(2010){Tiret}, {Salucci}, {Bernardi}, {Maraston}, \&
  {Pforr}}]{2010arXiv1009.5185T}
{Tiret}, O., {Salucci}, P., {Bernardi}, M., {Maraston}, C., \& {Pforr}, J.
  2010, ArXiv e-prints, 1009.5185

\bibitem[{{Trujillo} {et~al.}(2007){Trujillo}, {Conselice}, {Bundy}, {Cooper},
  {Eisenhardt}, \& {Ellis}}]{2007MNRAS.382..109T}
{Trujillo}, I., {Conselice}, C.~J., {Bundy}, K., {Cooper}, M.~C., {Eisenhardt},
  P., \& {Ellis}, R.~S. 2007, \mnras, 382, 109

\bibitem[{{van der Wel} {et~al.}(2009){van der Wel}, {Bell}, {van den Bosch},
  {Gallazzi}, \& {Rix}}]{2009ApJ...698.1232V}
{van der Wel}, A., {Bell}, E.~F., {van den Bosch}, F.~C., {Gallazzi}, A., \&
  {Rix}, H. 2009, \apj, 698, 1232

\bibitem[{{van Dokkum} {et~al.}(2008){van Dokkum}, {Franx}, {Kriek}, {Holden},
  {Illingworth}, {Magee}, {Bouwens}, {Marchesini}, {Quadri}, {Rudnick},
  {Taylor}, \& {Toft}}]{2008ApJ...677L...5V}
{van Dokkum}, P.~G., {Franx}, M., {Kriek}, M., {Holden}, B., {Illingworth},
  G.~D., {Magee}, D., {Bouwens}, R., {Marchesini}, D., {Quadri}, R., {Rudnick},
  G., {Taylor}, E.~N., \& {Toft}, S. 2008, \apjl, 677, L5

\bibitem[{{van Dokkum} {et~al.}(2010){van Dokkum}, {Whitaker}, {Brammer},
  {Franx}, {Kriek}, {Labb{\'e}}, {Marchesini}, {Quadri}, {Bezanson},
  {Illingworth}, {Muzzin}, {Rudnick}, {Tal}, \& {Wake}}]{2010ApJ...709.1018V}
{van Dokkum}, P.~G., {Whitaker}, K.~E., {Brammer}, G., {Franx}, M., {Kriek},
  M., {Labb{\'e}}, I., {Marchesini}, D., {Quadri}, R., {Bezanson}, R.,
  {Illingworth}, G.~D., {Muzzin}, A., {Rudnick}, G., {Tal}, T., \& {Wake}, D.
  2010, \apj, 709, 1018

\bibitem[{{Vitvitska} {et~al.}(2002){Vitvitska}, {Klypin}, {Kravtsov},
  {Wechsler}, {Primack}, \& {Bullock}}]{2002ApJ...581..799V}
{Vitvitska}, M., {Klypin}, A.~A., {Kravtsov}, A.~V., {Wechsler}, R.~H.,
  {Primack}, J.~R., \& {Bullock}, J.~S. 2002, \apj, 581, 799

\bibitem[{{Wiersma} {et~al.}(2010){Wiersma}, {Schaye}, {Dalla Vecchia},
  {Booth}, {Theuns}, \& {Aguirre}}]{2010arXiv1005.3921W}
{Wiersma}, R.~P.~C., {Schaye}, J., {Dalla Vecchia}, C., {Booth}, C.~M.,
  {Theuns}, T., \& {Aguirre}, A. 2010, ArXiv e-prints, 1005.3921

\bibitem[{{Zolotov} {et~al.}(2010){Zolotov}, {Willman}, {Brooks}, {Governato},
  {Hogg}, {Shen}, \& {Wadsley}}]{2010ApJ...721..738Z}
{Zolotov}, A., {Willman}, B., {Brooks}, A.~M., {Governato}, F., {Hogg}, D.~W.,
  {Shen}, S., \& {Wadsley}, J. 2010, \apj, 721, 738

\end{thebibliography}

\end{document}